%% file: Paper.tex
\pdfoutput=1
\newcommand*{\ATLASLATEXPATH}{latex/}
\documentclass[cernpreprint,UKenglish,texlive=2014,txfonts]{latex/atlasdoc}
\usepackage[subcaption, biblatex=true]{latex/atlaspackage}
\usepackage{float}
\usepackage{rotating}
\usepackage{multirow}
\newcommand{\hd}{\hphantom{0}}

\usepackage{latex/atlasbiblatex}

\usepackage{latex/atlascontribute}

\usepackage{latex/atlasphysics}
\graphicspath{{logos/}{figures/}}

\usepackage{Paper-defs}
\usepackage{latex/atlasmath}
\usepackage{latex/atlasunit}
\usepackage{latex/atlasbsm}
\usepackage{latex/atlasother}
\usepackage{latex/atlasxref}
\usepackage{latex/atlascontribute}
\usepackage{latex/atlasjournal}
\usepackage{latex/atlasmisc}
\usepackage{latex/atlaspackage}
\usepackage{latex/atlasprocess}

\input{Paper-metadata}

\hypersetup{pdftitle={ATLAS document},pdfauthor={The ATLAS Collaboration}}

\begin{document}
\maketitle

%-------------------------------------------------------------------------------
\input{introduction.tex}

\input{detector.tex}
\input{samples.tex}

\input{objects.tex}

\input{analysis-strategy.tex}

\input{bkg.tex}

\input{systematics.tex}

\input{results.tex}
\clearpage
\input{conclusions.tex}
\section*{Acknowledgements}
\input{Acknowledgements.tex}

\clearpage
  
%-------------------------------------------------------------------------------
\printbibliography

\newpage \input{atlas_authlist}

\end{document}

%% file: Paper-metadata.tex
\AtlasTitle{Search for new phenomena in a lepton plus high jet multiplicity final state with the ATLAS experiment using $\sqrt{s} = 13$~\tev\ proton--proton collision data}

\author{The ATLAS Collaboration}

\AtlasNote{SUSY-2016-11}

\PreprintIdNumber{CERN-EP-2017-053}

\AtlasJournalRef{JHEP09 (2017) 088}
\AtlasDOI{10.1007/JHEP09(2017)088}

\AtlasAbstract{%
A search for new phenomena in final states characterized by high jet multiplicity, an isolated lepton (electron or muon) and either zero or at least three $b$-tagged jets is presented. The search uses 
36.1~\ifb\ of $\sqrt{s}$ = 13~\tev\ proton--proton collision data collected by the ATLAS experiment at the Large Hadron Collider in 2015 and 2016. The dominant sources of background are estimated using 
parameterized extrapolations, based on observables at medium jet multiplicity, to predict the $b$-tagged jet multiplicity distribution at the higher jet multiplicities used in the search. No significant excess over the Standard Model expectation is observed and 95\% confidence-level limits are extracted constraining four simplified models of $R$-parity-violating supersymmetry that feature either 
gluino or top-squark pair production. The exclusion limits reach as high as 2.1 TeV in gluino mass and 1.2 TeV in top-squark mass in the models considered. In addition, an upper limit is set 
on the cross-section for Standard Model $t\bar{t}t\bar{t}$ production of 60~fb (6.5 $\times$ the Standard Model prediction) at 95\% confidence level. Finally, model-independent limits are set on the 
contribution from new phenomena to the signal-region yields.
}

%% file: introduction.tex
\section{Introduction}
\label{sec:intro}

The ATLAS experiment at the Large Hadron Collider (LHC)  has carried out a large number of searches for beyond the Standard Model (BSM) physics. These searches cover a broad range of different final-state particles and 
kinematics. However, one gap in the search coverage, as pointed out in Refs.~\cite{Lisanti2012, Evans2014}, is in final states with one or more leptons, many jets and no-or-little missing transverse 
momentum (the magnitude of which is denoted by \met ). Such a search is presented in this article, considering final states with an isolated lepton (electron or muon), at least eight to twelve jets 
(depending on the jet transverse momentum threshold), either zero or many $b$-tagged jets, and with no requirement on \met .

This search has potential sensitivity to a large number of BSM physics models. In this article, model-independent limits on the possible contribution of BSM physics to several single-bin  signal regions 
are presented. In addition, four $R$-parity-violating (RPV) supersymmetric (SUSY~\cite{Golfand:1971iw,Volkov:1973ix,Wess:1974tw,Wess:1974jb,Ferrara:1974pu,Salam:1974ig}) benchmark models are used to 
interpret the results. In this case, a multi-bin fit to the two-dimensional space of jet-multiplicity and $b$-tagged jet multiplicity is used to constrain the models.  The dominant Standard Model 
(SM) background arises from top-quark pair production and $W/Z$+jets production, with at least one lepton produced in the vector boson decay. The theoretical modelling of these backgrounds at high jet multiplicity suffers from large uncertainties, so they 
are estimated from the data by extrapolating the $b$-tagged jet multiplicity distribution extracted at moderate jet multiplicities to the high jet multiplicities of the search region.
Previous searches targeting similar RPV SUSY models have been carried out by the ATLAS and CMS collaborations~\cite{SUSY-2013-07,CMS-EXO-13-001,CMS-SUS-14-003,Aaboud:2017dmy}.

The result is also used to search for SM four-top-quark production. Previous searches for four-top-quark production were carried out by the ATLAS~\cite{EXOT-2013-18} and CMS~\cite{Sirunyan:2017tep} 
collaborations.

%% file: detector.tex
\section{ATLAS detector}
\label{sec:detector}

The ATLAS detector~\cite{PERF-2007-01} is a multipurpose detector with a forward-backward symmetric cylindrical
geometry and nearly 4$\pi$ coverage in solid angle.\footnote{ATLAS uses a right-handed coordinate system with its origin at the nominal interaction point in the centre of the detector.
The positive $x$-axis is defined by the direction from the interaction point to the centre of the LHC ring, with the positive $y$-axis 
pointing upwards, while the beam direction defines the $z$-axis. Cylindrical coordinates ($r$, $\phi$) are used in the transverse plane, $\phi$ 
being the azimuthal angle around the $z$-axis. The pseudorapidity $\eta$ is defined in terms of the polar angle $\theta$ by $\eta=-\ln\tan(\theta/2)$. 
The transverse momentum \pt\ is defined in the $x$--$y$ plane. 
Rapidity is defined as $y=0.5 \ln\left[(E + p_z )/(E - p_z )\right]$ where $E$ denotes the energy and $p_z$ is the component of the momentum along the beam direction.
} The inner detector (ID) tracking system consists of silicon pixel and microstrip detectors covering the pseudorapidity region |$\eta$| < 2.5, surrounded by a transition
radiation tracker which improves electron identification in the region |$\eta$| < 2.0. The innermost pixel
layer, the insertable B-layer~\cite{B-layerRef}, was added between Run 1 and Run 2 of the LHC, at a radius of 33 mm
around a new, narrower and thinner, beam pipe. The ID is surrounded by a thin superconducting solenoid
providing an axial 2 T magnetic field and by a fine-granularity lead/liquid-argon (LAr) electromagnetic
calorimeter covering |$\eta$| < 3.2. A steel/scintillator-tile calorimeter provides hadronic calorimetry in the
central pseudorapidity range (|$\eta$| < 1.7). The endcap and forward regions (1.5 < |$\eta$| < 4.9) of the hadronic
calorimeter are made of LAr active layers with either copper or tungsten as the absorber material. A 
muon spectrometer with an air-core toroidal magnet system surrounds the calorimeters. Three layers of
high-precision tracking chambers provide coverage in the range |$\eta$| < 2.7, while dedicated chambers
allow triggering in the region |$\eta$| < 2.4.

The ATLAS trigger system~\cite{Aaboud:2016leb}
 consists of two levels; the first level is a hardware-based system, while the second is a software-based system 
called the High-Level Trigger. 

%% file: samples.tex
\section{Data and simulated event samples}
\label{sec:samples}

\subsection{Data sample}
\label{sec:datasample}
After applying beam, detector and data-quality criteria, the data sample analysed comprises 36.1~\ifb\ of $\sqrt{s}$ = 13~\tev\ proton--proton ($pp$) collision data (3.2~\ifb\ collected in 
2015 and 32.9~\ifb\ collected in 2016) with a minimum $pp$ bunch spacing of 25 ns. 
In this data set, the mean number of $pp$ interactions per proton-bunch crossing (pile-up) is $\left<\mu\right>$ = 23.7. 
The luminosity and its uncertainty of $3.2\%$ are derived following a methodology similar to that detailed in Ref.~\cite{DAPR-2013-01} from a preliminary calibration of the luminosity 
scale using a pair of $x$--$y$ beam separation scans performed in August 2015 and June 2016.

Events are recorded online using a single-electron or single-muon trigger with thresholds that give a constant efficiency as a function of lepton-\pT\ of $\approx$90\% ($\approx$80\%) for electrons (muons) 
for the event selection used. For the determination of the multi-jet background, alternative lepton triggers, using less stringent lepton isolation requirements with respect to the nominal ones, are considered, as 
discussed in Section~\ref{sec:bkg}. 
Single-photon and multi-jet triggers are also employed to select data samples used in the validation of the background estimation technique. 

\subsection{Simulated event samples}
\label{sec:simulatedsamples}
Samples of Monte Carlo (MC) simulated events are used to model the signal and to validate the background estimation procedure for the dominant background contributions. 
In addition, simulated events are used to model the sub-dominant 
background processes. The response of the detector to particles is modelled with a full ATLAS detector simulation~\cite{SOFT-2010-01} based on {\scshape Geant4}~\cite{Agostinelli:2002hh}, or with a fast 
simulation based on a parameterization of the response of the ATLAS electromagnetic and hadronic calorimeters~\cite{ATL-PHYS-PUB-2010-013} and on {\scshape Geant4} elsewhere.
All simulated events are overlaid with pile-up collisions simulated with the soft strong interaction processes of \pythia 8.186~\cite{Sjostrand:2007gs} using the A2 set of tunable parameters 
(tune)~\cite{ATLAS:2011krm} and the MSTW2008LO~\cite{Sherstnev:2007nd} parton distribution function (PDF) set. 
The simulated events are reconstructed in the same way as the data, and are reweighted so that the distribution of the expected number of collisions per bunch crossing matches the one in the data. 

For all MC samples used, except those produced by the \sherpa event generator, the \textsc{EvtGen} 1.2.0 program~\cite{Lange:2001uf} is used to model the properties of bottom and charm hadron decays.

\subsubsection{Simulated signal events}
\label{sec:signal}
Simulated signal events from four SUSY benchmark models are used to guide the analysis selections and to estimate the expected signal yields for different signal-mass hypotheses used to interpret 
the analysis results. In all models, the RPV couplings and the SUSY particle masses are chosen to ensure prompt decays of the SUSY particles. 
Diagrams of the first three benchmark simplified models, which involve gluino pair production, are shown in Figures~\ref{fig:diagram}(a), 1(b), and 1(c). 
In the first model, each gluino decays via a virtual top squark to two top quarks 
and the lightest neutralino (\ninoone) which is the lightest supersymmetric particle (LSP). The \ninoone decays to three light quarks ($\ninoone \to u d s$) via the RPV coupling $\lambda_{112}''$.
For this model, \ninoone\ masses below 10~\gev\ are not considered in order to avoid the effect of the limited phase space in the \ninoone\ decay. 
In the second model, each gluino decays to a top quark and a top squark LSP, with the top squark decaying to an $s$-quark and a $b$-quark via a non-zero $\lambda_{323}''$ RPV coupling.\footnote{The same final state can be produced by requiring a non-zero $\lambda_{313}''$ RPV coupling, however the minimal flavour violation hypothesis~\cite{DAmbrosio:2002ex} favours a large $\lambda_{323}''$ coupling~\cite{Csaki:2011ge}.} 
The third model involves the gluino decaying to two first or second generation quarks ($q \equiv (u,d,s,c)$) and the \ninoone\ LSP, which then decays to two additional first or second generation quarks and a charged lepton or a neutrino 
($\ninoone \to q \bar{q}' \ell$ or  $\ninoone \to q \bar q  \nu$, labelled as $\ninoone \to q \bar q \ell / \nu$). 
The decay proceeds via a $\lambda'$  RPV coupling, where each RPV decay can produce any of the four first- and second-generation leptons ($e^{\pm},\mu^{\pm},\nu_e,\nu_\mu$) with equal probability. 
For this model, \ninoone\ masses below 50~\gev\ are not considered. 

The fourth scenario considered involves right-handed top-squark pair production with the top squark decaying to a bino or higgsino LSP and a top or bottom quark.
The LSP decays through the non-zero RPV coupling $\lambda_{323}'' \approx \mathcal{O}(10^{-2}$--$10^{-1})$, with the value chosen to ensure prompt decays for the particle masses considered\footnote{LSP masses below 200~\gev\ are not considered as in this case non-prompt RPV decays can occur.} and to avoid more complex patterns of RPV decays that are not considered here. 
Figure~\ref{fig:diagram}(d) shows the production and possible decays considered. 
The different decay modes depend on the nature of the LSP and have a small dependence on the top-squark mass, 
with the top squark decaying as: $\stop \to t \ninoone$ for a bino-like LSP  and as $\stop \to t \ninotwo$ ($\approx$25\%), $\stop \to t \ninoone$ ($\approx$25\%),  $\stop \to b \chinoonep$ ($\approx$50\%) for higgsino-like LSPs. With the chosen model parameters, the electroweakinos decay as $\tilde\chi^0_{1/2} \to t b s$ or $\tilde\chi^\pm_1 \to b b s$. The search results are 
interpreted in this model, with the assumption of either a pure higgsino ($\tilde{H}$) or pure bino ($\tilde{B}$) LSP. In the case of a wino LSP, the search has no sensitivity as the top squark decays directly as $\stop \to \bar b \bar s$ with no 
leptons produced in the final state.\footnote{For this case, a dedicated ATLAS search~\cite{SUSY-2015-05} excludes top-squark masses up to 315~\GeV .}

Event samples for the first signal model ($\gluino \to t \bar t \ninoone \to t \bar t u d s$) are produced using the \textsc{Herwig++ 2.7.1}~\cite{Corcella:2000bw} event generator with the 
\textsc{cteq6l1}~\cite{Pumplin:2002vw} PDF set, and the UEEE5 tune~\cite{Gieseke:2012ft}. For the other three models, the MG5\_aMC@NLO v2.3.3~\cite{Alwall:2014hca} event generator interfaced to \pythia 8.210 is used. 
For these cases, signal events are produced with either one ($\gluino\to \bar t \stop \to \bar t \bar b \bar s$ model) or two ($\gluino \to q \bar q \ninoone \to q \bar q q \bar q \ell / \nu$ and $\stop \to t \tilde{H} / \tilde{B}$ models) additional partons in the matrix element and using the A14~\cite{ATL-PHYS-PUB-2014-021} tune. The parton luminosities are provided by the 
{\scshape NNPDF23LO}~\cite{Ball:2012cx} PDF set.
 
Signal cross-sections are calculated to next-to-leading order in the strong coupling constant, adding the resummation of soft-gluon emission at next-to-leading-logarithmic accuracy 
(NLO+NLL)~\cite{Beenakker:1996ch,Kulesza:2008jb,Kulesza:2009kq,Beenakker:2009ha,Beenakker:2011fu}. The nominal cross-section and its uncertainty are taken from an envelope of cross-section 
predictions using different PDF sets as well as different factorization and renormalization scales, as described in Ref.~\cite{Borschensky:2014cia}.

The analysis is also used to search for SM four-top-quark production. In this case, the $t\bar{t}t\bar{t}$ sample is generated with the \madgraph 2.2.2 event generator interfaced to \pythia8.186 using the {\scshape NNPDF23LO} PDF set and the A14 tune. 

\begin{figure}[htb]
\centering
\begin{subfigure}[t]{0.45\textwidth}
\includegraphics[width=\textwidth]{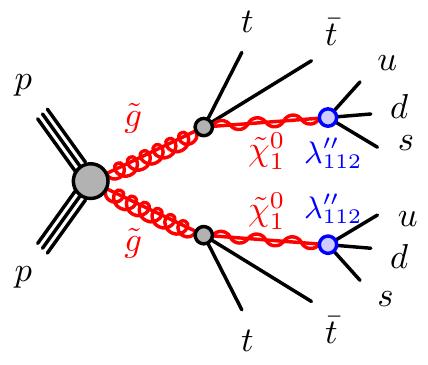}
\caption{}\label{fig:model1}\end{subfigure}
\begin{subfigure}[t]{0.45\textwidth}
\includegraphics[width=\textwidth]{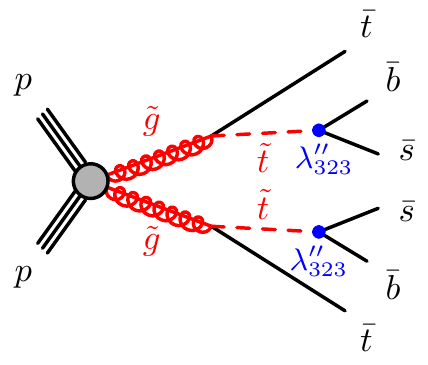}
\caption{}\label{fig:model2}\end{subfigure}
\begin{subfigure}[t]{0.45\textwidth}
\includegraphics[width=\textwidth]{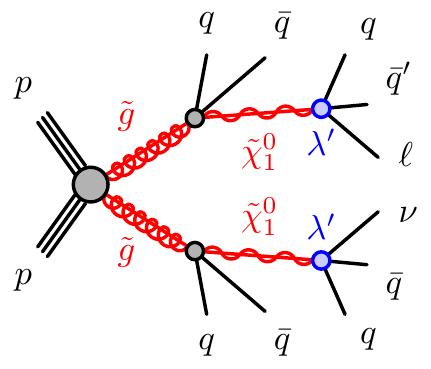}
\caption{}\label{fig:model3}\end{subfigure}
\begin{subfigure}[t]{0.45\textwidth}
\includegraphics[width=\textwidth]{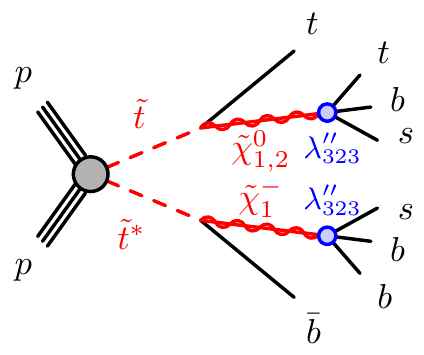}
\caption{}\label{fig:model4}\end{subfigure}
\caption{ Diagrams of the four simplified signal benchmark models considered. The first three models involve pair production of gluinos with each gluino decaying as (a) $\gluino \to t \bar t \ninoone \to t \bar t u d s$,  (b) $\gluino \to  \bar t \stop \to \bar{t} \bar{b} \bar{s}$,  (c)  $\gluino \to q \bar q \ninoone \to q \bar q q \bar q \ell/\nu$. The fourth model (d) involves pair production of top squarks with the decay $\stop \to t \tilde\chi^0_{1/2}$ or $\stop \to b \chinoonep$ and with the LSP decays $\tilde\chi^0_{1/2} \to t b s$ or $\chinoonep \to \bar b \bar b \bar s$; the specific decay depends on the nature of the LSP. In all signal scenarios, anti-squarks decay into the charge-conjugate final states of those indicated for the corresponding squarks, and each gluino decays with equal probabilities into the given final state or its charge conjugate. }
\label{fig:diagram}
\end{figure}

\subsubsection{Simulated background events}
\label{sec:bkg-samples}
The dominant backgrounds from top-quark pair production and $W/Z$+jets production are estimated from the data as described in Section~\ref{sec:bkg}, whereas the expected yields for minor backgrounds are 
taken from MC simulation. In addition, the background estimation procedure is validated with simulated events, and some of the systematic uncertainties are estimated using simulated event samples.
The samples used are shown in Table~\ref{tab:MC} and more details of the event generator configurations can be found in Refs.~\cite{ATL-PHYS-PUB-2016-004,ATL-PHYS-PUB-2016-003,ATL-PHYS-PUB-2016-002,ATL-PHYS-PUB-2016-005}.

\begin{table*}[ht]
  \begin{center}
  \scriptsize
  \begin{tabular}{l l c c c c}
  \hline\hline
  Physics process & Event & Parton-shower & Cross-section & PDF set & Tune\\
  &generator &modelling & normalization & & \\
  \hline\hline
  $W(\rightarrow \ell\nu)$ + jets              & \sherpa 2.2.1~\cite{sherpa2}    & \sherpa 2.2.1     & NNLO~\cite{Catani:2009sm}  & NLO CT10~\cite{Lai:2010vv}    & \sherpa default \\
  $W(\rightarrow \ell\nu)$ + jets (*)              &  \madgraph 2.2.2  & \pythia8.186   & NNLO  & NNPDF2.3LO & A14    \\
  $W(\rightarrow \ell\nu)$ + jets (*)              &  \alpgen v2.14~\cite{Mangano:2002ea}  & \pythia6.426~\cite{Sjostrand:2006za}   & NNLO  & CTEQ6L1 & \scshape{Perugia2012}~\cite{Skands:2010ak}    \\
  $Z/\gamma^{*}(\rightarrow \ell \ell)$ + jets & \sherpa 2.2.1   & \sherpa 2.2.1       & NNLO~\cite{Catani:2009sm}  & NLO CT10         & \sherpa default\\
  $t\bar{t}$ + jets             & {\scshape powheg-box} v2~\cite{Alioli:2010xd} & \pythia6.428  & NNLO+NNLL~\cite{Czakon:2013goa,Czakon:2012pz,Czakon:2012zr,Baernreuther:2012ws,Cacciari:2011hy,Czakon:2011xx}  & NLO CT10  &\scshape{Perugia2012}\\
  $t\bar{t}$ + jets (*)              & \madgraph 2.2.2 & \pythia8.186  & NNLO+NNLL  & NNPDF2.3LO & A14    \\
  
  Single-top              &&&&\\
  ($t$-channel)           & {\scshape powheg-box} v1 & \pythia6.428 & NNLO+NNLL~\cite{Kidonakis:2011wy}  & NLO CT10f4   & \scshape{Perugia2012}\\
  Single-top              &&&&\\
  ($s$- and $Wt$-channel) & {\scshape powheg-box} v2 & \pythia6.428  & NNLO+NNLL~\cite{Kidonakis:2010ux,Kidonakis:2010tc}  & NLO CT10   & \scshape{Perugia2012}\\ 
  
  $t\bar{t}+W/Z/WW$       &  \madgraph 2.2.2  & \pythia8.186   & NLO~\cite{Alwall:2014hca}  & NNPDF2.3LO & A14    \\
  
  $WW$, $WZ$ and $ZZ$     &  \sherpa 2.2.1  & \sherpa 2.2.1      & NLO  & NLO CT10   & \sherpa default \\
  $t\bar{t}H$     & \madgraph 2.3.2  &  \pythia8.186     & NLO~\cite{Dittmaier:2011ti}  &  NNPDF2.3LO & A14  \\
 $t\bar{t}t\bar{t}$     & \madgraph 2.2.2  & \pythia8.186   & NLO~\cite{Alwall:2014hca}  & NNPDF2.3LO & A14    \\

  \hline\hline
  \end{tabular}
  \caption{Simulated background event samples:
             the corresponding event generator, parton-shower modelling, cross-section normalization,
             PDF set and underlying-event tune are shown. The samples marked with (*) are alternative samples used to validate the background estimation method.
    }
  
  \label{tab:MC}
  \end{center}
  \end{table*}

%% file: objects.tex
%-------------------------------------------------------------------------------
\section{Event reconstruction}
\label{sec:objects}
For a given event, primary vertex candidates are required to be consistent with the luminous region and to have at least two associated tracks with $\pt > 400$~\MeV. The vertex with the largest  
$\sum \pt^2$ of the associated tracks is chosen as the primary vertex of the event.

Jet candidates are reconstructed using the anti-$k_{t}$ jet clustering algorithm~\cite{Cacciari:2008gp,Cacciari:2005hq} with a radius parameter of $0.4$ starting from energy deposits in clusters of calorimeter 
cells \cite{PERF-2014-07}. The jets are corrected for energy deposits from pile-up collisions using the method suggested in Ref.~\cite{Cacciari:2007fd} and calibrated with ATLAS data
in Ref.~\cite{ATLAS-CONF-2013-083}: a contribution equal to the product of the jet area and the median energy density of the event is subtracted from the jet energy.
Further corrections derived from MC simulation and data are used to calibrate on average the energies of jets to the scale of their constituent particles~\cite{PERF-2016-04}. 
In the search, three jet \pt thresholds of $40~\GeV$, $60~\GeV$ and $80~\GeV$ are used, with all jets required to have $|\eta|<2.4$. 
To minimize the contribution from jets arising from pile-up interactions, the selected jets must satisfy a loose jet vertex tagger (JVT) requirement~\cite{ATLAS-CONF-2014-018}, where JVT is an algorithm 
that uses tracking and primary vertex information to determine if a given jet originates from the primary vertex. The chosen working point has an efficiency of 94\% at a jet \pT\ of 40~\GeV\ and is 
nearly fully efficient above 60~\GeV\ for jets originating from the hard parton--parton scatter. This selection reduces the number of jets originating from, or heavily contaminated by, pile-up interactions, to a negligible level. 
Events with jet candidates originating from detector noise or non-collision background are rejected if any of the jet candidates satisfy the `LooseBad' quality criteria, described in Ref.~\cite{ATLAS-CONF-2015-029}. 
The coverage of the calorimeter and the jet reconstruction techniques allow high-jet-multiplicity final states to be reconstructed efficiently. For example, 12 jets take up only about one fifth of the available solid angle. 

Jets containing a $b$-hadron ($b$-jets) are identified by a multivariate algorithm using information about the impact parameters of ID tracks matched to the jet, the presence of displaced secondary 
vertices, and the reconstructed flight paths of $b$- and $c$-hadrons inside the jet~\cite{Aad:2015ydr}. The operating point used corresponds to an efficiency of 78\%\ in simulated $t\bar{t}$ events, 
along with a rejection factor of approximately 110 for jets induced by gluons or light quarks and of 8 for charm jets~\cite{ATL-PHYS-PUB-2016-012}, and is configured to give a constant $b$-tagging efficiency as a function of jet \pT . 

Since there is no requirement on \met\ or any \met\ derived quantity 
the search is particularly sensitive to fake or non-prompt leptons in multi-jet events. In order to suppress this background to an acceptable level, stringent lepton identification and isolation 
requirements are used.

Muon candidates are formed by combining information from the muon spectrometer and the ID and must satisfy the `Medium' quality criteria described in Ref.~\cite{MuonPerfRun2}. 
They are required to have $\pt > 30~\GeV$ and $|\eta| <2.4$. Furthermore, they must satisfy requirements on the significance of the transverse impact parameter with respect to the primary vertex, 
$|d_0^{\mathrm{PV}}|/\sigma(d_0^{\mathrm{PV}}) <$ 3, the longitudinal impact parameter with respect to the primary vertex,  $|z_0^{\mathrm{PV}} \mathrm{sin}(\theta)|<$ 0.5~mm, and the `Gradient' 
isolation requirements, described in Ref.~\cite{MuonPerfRun2}, relying on a set of $\eta$- and $\pt$-dependent criteria based on tracking- and calorimeter-related variables. 

Electron candidates are reconstructed from isolated energy deposits in the electromagnetic calorimeter matched to ID tracks and are required to have $\pt > 30~\GeV$, $|\eta| < 2.47$, and to satisfy the 
`Tight' likelihood-based identification criteria described in Ref.~\cite{ATLAS-CONF-2016-024}. Electron candidates that fall in the transition region between the barrel and endcap calorimeters 
($1.37< |\eta| < 1.52$) are rejected. They are also required to have $|d_0^{\mathrm{PV}}|/\sigma(d_0^{\mathrm{PV}}) <$ 5, $|z_0^{\mathrm{PV}} \mathrm{sin}(\theta)|<$ 0.5~mm, and to satisfy 
isolation requirements described in Ref.~\cite{ATLAS-CONF-2016-024}. % similar to those applied to muon candidates. 

An overlap removal procedure is carried out to resolve ambiguities 
between candidate jets (with $\pt > 20~\GeV$) and baseline leptons\footnote{Baseline leptons are reconstructed as described above, but with a looser \pt\ requirement ($\pt > 10~\GeV$), no 
isolation or impact parameter requirements, and, in the case of electrons, the `Loose' lepton identification  criteria~\cite{ATL-PHYS-PUB-2015-041}.} as follows: 
first, any non-$b$-tagged jet candidate\footnote{In this case, a $b$-tagging working point corresponding to an efficiency of identifying $b$-jets in a simulated $t\bar{t}$ sample of 85\%\ is used.} 
lying within an angular distance $\Delta R\equiv\sqrt{(\Delta y)^2+(\Delta\phi)^2}=0.2$ of a baseline electron is discarded. Furthermore, non-$b$-tagged jets within  $\Delta R = 0.4$ of baseline muons are 
removed if the number of tracks associated with the jet is less than three or the ratio of muon \pT\ to jet \pT\ is greater than 0.5. Finally, any baseline lepton candidate remaining within a 
distance $\Delta R = 0.4$ of any surviving jet candidate is discarded.

Corrections derived from data control samples are applied to account for differences between data and simulation for the lepton trigger, reconstruction, identification and isolation efficiencies, the 
lepton momentum/energy scale and resolution~\cite{ATLAS-CONF-2016-024,MuonPerfRun2}, and for the efficiency and mis-tag rate of the $b$-tagging algorithm~\cite{Aad:2015ydr}.

%% file: analysis-strategy.tex
\section{Event selection and analysis strategy}
\label{sec:anal-strategy}

Events are selected online using a single-electron or single-muon trigger. For the analysis selection, at least one electron or muon, matched to the trigger lepton, is required in the event. The analysis is carried out with three sets of 
jet \pT thresholds to provide sensitivity to a broad range of possible signals. These thresholds are applied to all jets in the event and are \pT $= 40~\GeV, 60~\GeV$, and $80~\GeV$. 
The jet multiplicity is binned from a minimum of five jets to a maximum number that depends on the \pT\ threshold. The last bin is inclusive, so that it also includes all events with more jets than the bin number. This bin corresponds to 12 or more jets for the $40$~\GeV\ requirement, and 10 or more jets for the $60$~\GeV\ and $80$~\GeV\ thresholds. 
There are five bins in the $b$-tagged jet multiplicity 
(exclusive bins from zero to three with an additional inclusive four-or-more bin). In this article, the notation $N^{\text{process}}_{\text{$j$},\text{$b$}}$ is used to denote the number of events 
predicted by the background fit model, with $j$ jets and $b$ $b$-tagged jets for a given process, e.g. $N^{\text{$t\bar{t}$+jets}}_{\text{$j$},\text{$b$}}$ for $t\bar{t}$+jets events. The number of 
events summed over all $b$-tag multiplicity bins for a given number of jets is denoted by $N^{\text{process}}_{\text{$j$}}$, and is also referred to as a jet slice.

For probing a specific BSM model, all of these bins in data are simultaneously fit to constrain the model, in what is labeled a model-dependent fit. In the search for a hypothetical BSM signal, 
dedicated signal regions (SRs) are defined which could be populated by a signal, and where the SM contribution is expected to be small. The background in these SRs is estimated from 
a fit in which some of the bins can be excluded to limit the effect of signal contamination biasing the background estimate; this set-up is labeled a model-independent fit. More details of the SR definitions are given in Section~\ref{subsec:bkd_validation}.

\begin{figure}[htbp]
\centering
\includegraphics[width=.7\textwidth]{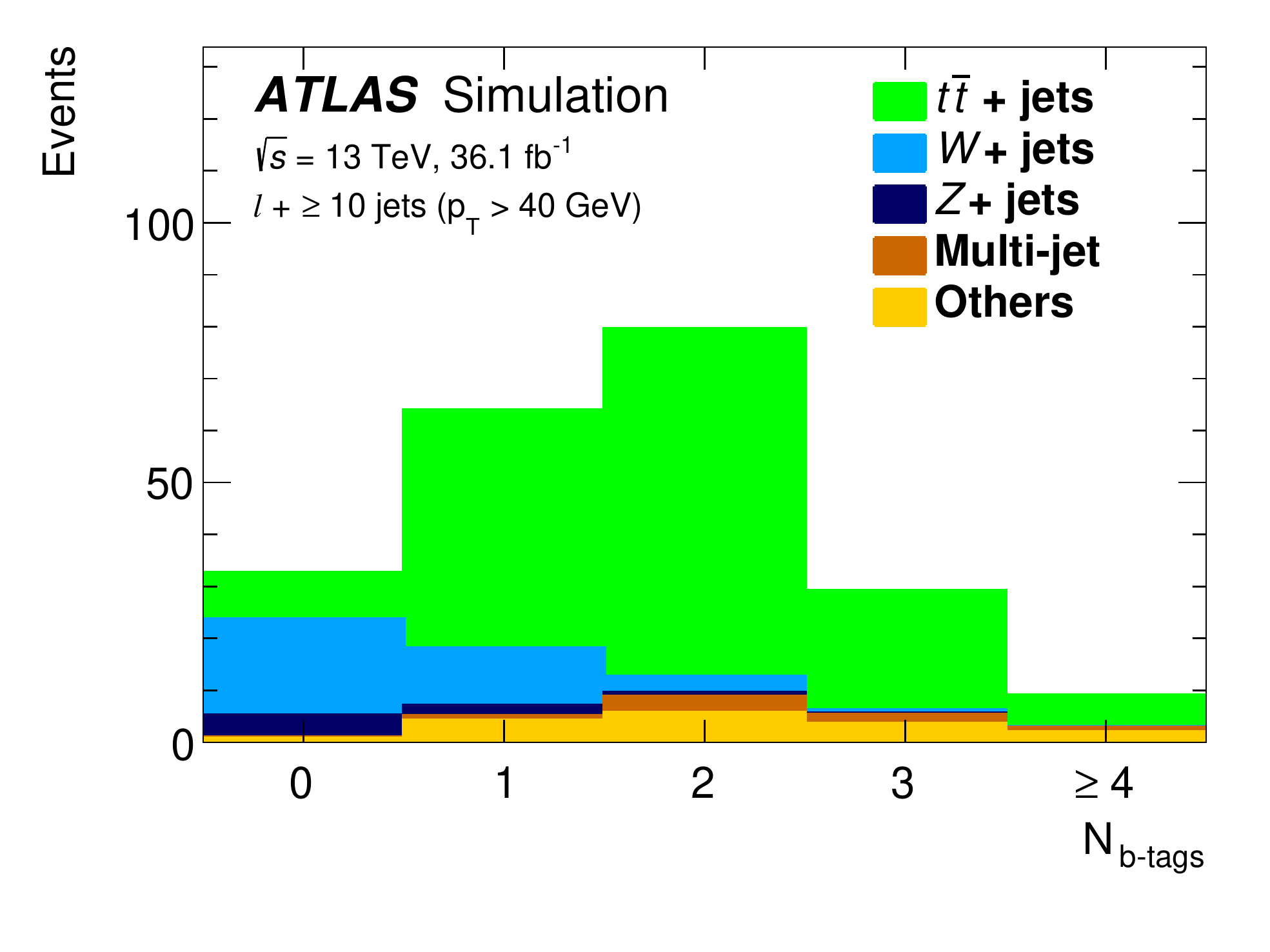}
\caption{The expected background from MC simulation in the different $b$-tag bins, with a selection of at least ten jets  (with \pT\ $> 40$~\GeV ).}
\label{fig:mcBkg}
\end{figure}

An example of the expected background contributions from MC simulation for the different $b$-tag bins, with a selection of at least ten jets, can be seen in Figure~\ref{fig:mcBkg}. This figure shows 
that  the background in the zero $b$-tag bin is dominated by $W/Z$+jets and $t\bar{t}$+jets, whereas in the other $b$-tag bins it is dominated by $t\bar{t}$+jets. The contribution from other 
processes is very small in all bins.

The estimation of the dominant background processes of $t\bar{t}$+jets and $W/Z$+jets production is carried out using a combined fit to the jet and $b$-tagged jet multiplicity bins described above. 
For these backgrounds, the normalization per jet slice is derived using  parameterized extrapolations from lower jet multiplicities. The $b$-tag multiplicity shape per jet slice is taken from simulation 
for the $W/Z$+jets background, whereas for the $t\bar{t}$+jets background it is predicted from the data using a parameterized extrapolation based on observables at medium jet multiplicities.
A separate likelihood fit is carried out for each jet \pT\ threshold, with the fit parameters of the background model determined separately in each fit. 
The assumptions used in the parameterization are validated using data and MC simulation.  
Regarding the model-independent results, it is to be noted that possible signal leakage to the control regions can produce a bias in the background estimation. 
Such limits have been hence obtained assuming negligible signal contributions to events with five, six or seven jets. 
Signal processes with final states that the search is targeting, generally have negligible leakage into these jet slices, as is the case for the benchmark models considered.

%% file: bkg.tex
\section{Background estimation}
\label{sec:bkg}

\subsection{$W/Z$+jets}
\label{sec:bkd_vjets}
A partially data-driven approach is used to estimate the $W/Z$+jets background. 
Since the selected $W/Z$+jets background events usually have no $b$-jets, the shapes of the $b$-tag multiplicity distributions are taken from simulated events, whereas the normalization in each jet slice is derived from the data. 
The estimate of the normalization relies on assuming a functional form to describe the evolution of the number of $W/Z$+jets events as a function of the jet multiplicity, 
$r(j) \equiv N^{\text{$W/Z$+jets}}_{j+1}/N^{\text{$W/Z$+jets}}_{j}$. 

Above a certain number of jets, $r(j)$ can be assumed to be constant, implying a fixed probability of additional jet radiation, referred to as 
``staircase scaling''~\cite{Ellis:1985vn,Berends:196850,Giele1990,Gerwick2012}. This behaviour has been observed by the ATLAS~\cite{STDM-2012-04,Aaboud:2017hbk} and CMS~\cite{Chatrchyan2012} 
collaborations. For lower jet multiplicities, a different scaling is expected with $r(j) = k / (j + 1)$ where $k$ is a constant, referred to as ``Poisson scaling''~\cite{Gerwick2012}.\footnote{The 
transition between these scaling behaviours depends on the jet kinematic selections.}

For the kinematic phase space relevant for this search, a combination of the two scalings is found to describe the data in dedicated validation regions (described later in this section), 
as well as in simulated $W/Z$+jets event samples with an integrated luminosity much larger than the one of the data. 
This combined scaling is parameterized as 
\begin{equation}
r(j) = c_0 +  c_1 / (j + 1), 
\label{eqn:scaling}
\end{equation}
where $c_0$ and $c_1$ are constants that are extracted from the data. Studies using simulated event samples, both at generator level and after event reconstruction, demonstrate that the flexibility of this parameterization is also able to absorb reconstruction effects 
related to the decrease in event reconstruction efficiency with increasing jet multiplicity, which are mainly due to the lepton--jet overlap and lepton isolation requirements.

The number of $W$+jets or $Z$+jets events with different jet and $b$-jet multiplicities, $N^{\text{$W/Z$+jets}}_{j,b}$, is then parameterized as follows:
\begin{equation*}
N^{\text{$W/Z$+jets}}_{j,\text{$b$}} = \frac{MC^{\text{$W/Z$+jets}}_{j,\text{$b$}}}{MC^{\text{$W/Z$+jets}}_{j}} \cdot N^{\text{$W/Z$+jets}}_{5} \cdot \prod\limits_{j'=5}^{j'=j-1} r(j') ,
\end{equation*}

where $MC^{\text{$W/Z$+jets}}_{j,b}$ and $MC^{\text{$W/Z$+jets}}_{j}$ are the predicted numbers of $W/Z$ + $j$ jets events with $b$ $b$-tags and inclusive in $b$-tags, respectively, 
both taken from MC simulation, and $N^{\text{$W/Z$+jets}}_{5}$ is the absolute normalization in five-jet events. The term 
$N^{\text{$W/Z$+jets}}_{5} \cdot \prod\nolimits_{j'=5}^{j'=j-1} r(j')$ gives the number of $b$-tag inclusive events in jet slice $j$, and the ratio 
$MC^{\text{$W/Z$+jets}}_{j,\text{$b$}} / MC^{\text{$W/Z$+jets}}_{j}$ is the fraction of $b$ $b$-tagged events in this jet slice. 
The four parameters $N^{\text{$W$+jets}}_{5}, N^{\text{$Z$+jets}}_{5}$, $c_0$, and $c_1$ are left floating in the fit and are therefore extracted from the data along with the other background contributions.

Due to different $b$-tagged-jet multiplicity distributions in $W$+jets and $Z$+jets events, the $b$-tag distribution is modelled separately for the two processes. 
The normalization and scaling parameters $N^{\text{$W/Z$+jets}}_{5}$, $c_0$, and $c_1$ are determined using control regions with five, six or seven jets 
and zero $b$-tags. For the $Z$+jets background determination, the control regions are defined by selecting events with two oppositely charged same-flavour leptons fulfilling an invariant-mass requirement around 
the $Z$-boson mass (81$\le m_{\ell\ell} \le$ 101~\GeV ), as well as the requirement of exactly five, exactly six or exactly seven jets, and zero $b$-tags. The determination of the $W$+jets background relies on control 
regions containing the remaining events with exactly five, six or seven jets, and zero $b$-tags, which, for each jet multiplicity, are split according to the electric charge of the highest-\pT\ lepton. The 
expected charge asymmetry in $W$+jets events is taken from MC simulation separately for five-jet, six-jet and seven-jet events and used to constrain the $W$+jets normalization from the data using 
these control regions. Although all parameters are determined in a global likelihood fit, the most powerful constraint on the absolute normalization comes from the five-jet control regions, and the 
dominant constraints on the $c_0$ and $c_1$  parameters originate from the combination of the five-jet, six-jet and seven-jet control regions. The contamination by $\ttbar$ events in the $Z$+jets two-lepton 
control regions is negligible, whereas in the control regions used to estimate the $W$+jets normalization it is significant and is discussed in Section~\ref{subsec:bkd_tt}. Once the $W$+jets and $Z$+jets
 backgrounds are normalized, they are extrapolated to higher jet multiplicities using the same common scaling function $r(j)$. While independent scalings could be used, tests in data show no significant difference and therefore a common 
function is used.

The jet-scaling assumption is validated in data using $\gamma$+jets and multi-jet events, and simulated $W$+jets and $Z$+jets samples are also found to be consistent with this assumption. The $\gamma$+jets 
events are selected using a photon trigger, and an isolated photon~\cite{ATLAS-CONF-2012-123} with $\pT > 145~\GeV$ is required in the event selection, whereas the multi-jet events are selected 
using prescaled and unprescaled multi-jet triggers. In both cases, selections are applied to ensure these control regions probe a kinematic phase-space region similar to the one relevant for the analysis. 

\begin{figure}[htb]
\centering
\includegraphics[width=.7\textwidth]{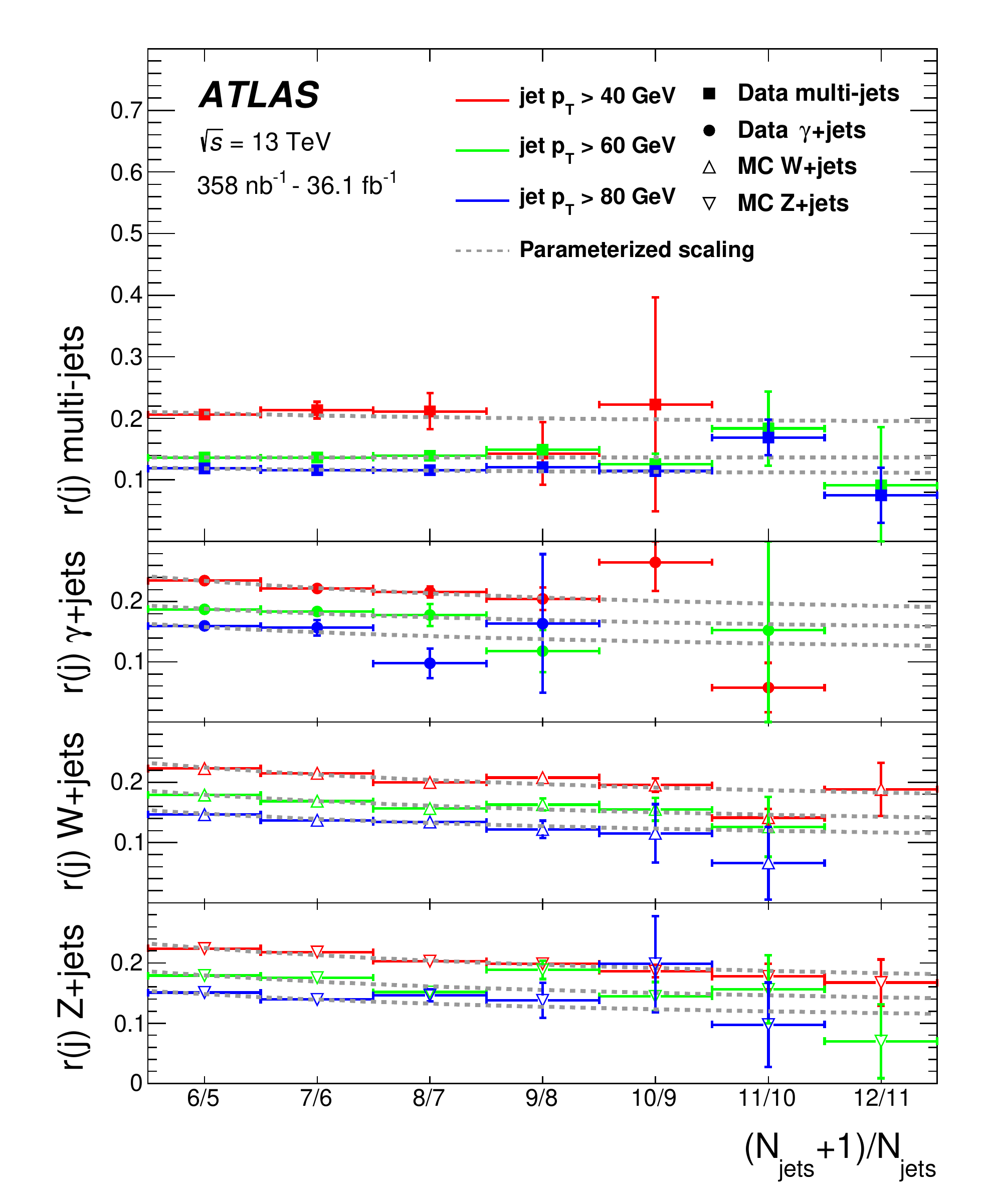}
\caption{ 
The ratio of the number of events with $(j+1)$ jets to the number with $j$ jets for various processes used to validate the jet-scaling parameterization. Each panel shows
the ratio for data or MC simulation with the fitted parameterization overlaid as a line. In the case of pure ``staircase scaling'', the shown ratio would be a constant.
For the multi-jet data points, the 40~\GeV\ jet \pT\ selection uses a prescaled trigger corresponding to an integrated luminosity of 358 nb$^{-1}$; all other selections use unprescaled triggers corresponding to the full data set. The uncertainties shown are statistical. 
}
\label{fig:v+jets}
\end{figure}

Figure~\ref{fig:v+jets} shows the $r(j)$ ratio 
for various processes used to validate the jet-scaling parameterization. Each panel shows 
the ratio for data or MC simulation with the fitted parameterization overlaid as a line. In the case of pure ``staircase scaling'', the shown ratio would be a constant. %, which is nearly the case in many cases. Javier: this would go against our logic that we do need to the Poisson part right?

Since the last jet-multiplicity bin used in the analysis is inclusive in the number of jets, the $W/Z$+jets background model is used to predict this by iterating to higher jet multiplicities and summing the contribution for each jet multiplicity above the maximum used in the analysis, and therefore gives the correct inclusive yield in this bin.

%%%%%%%%%%%%%%%%%%%%%%%%%%%%% ttbar + jets
\subsection{$\ttbar$+jets}
\label{subsec:bkd_tt}
\begin{figure}[htb]
\centering
\includegraphics[width=.7\textwidth]{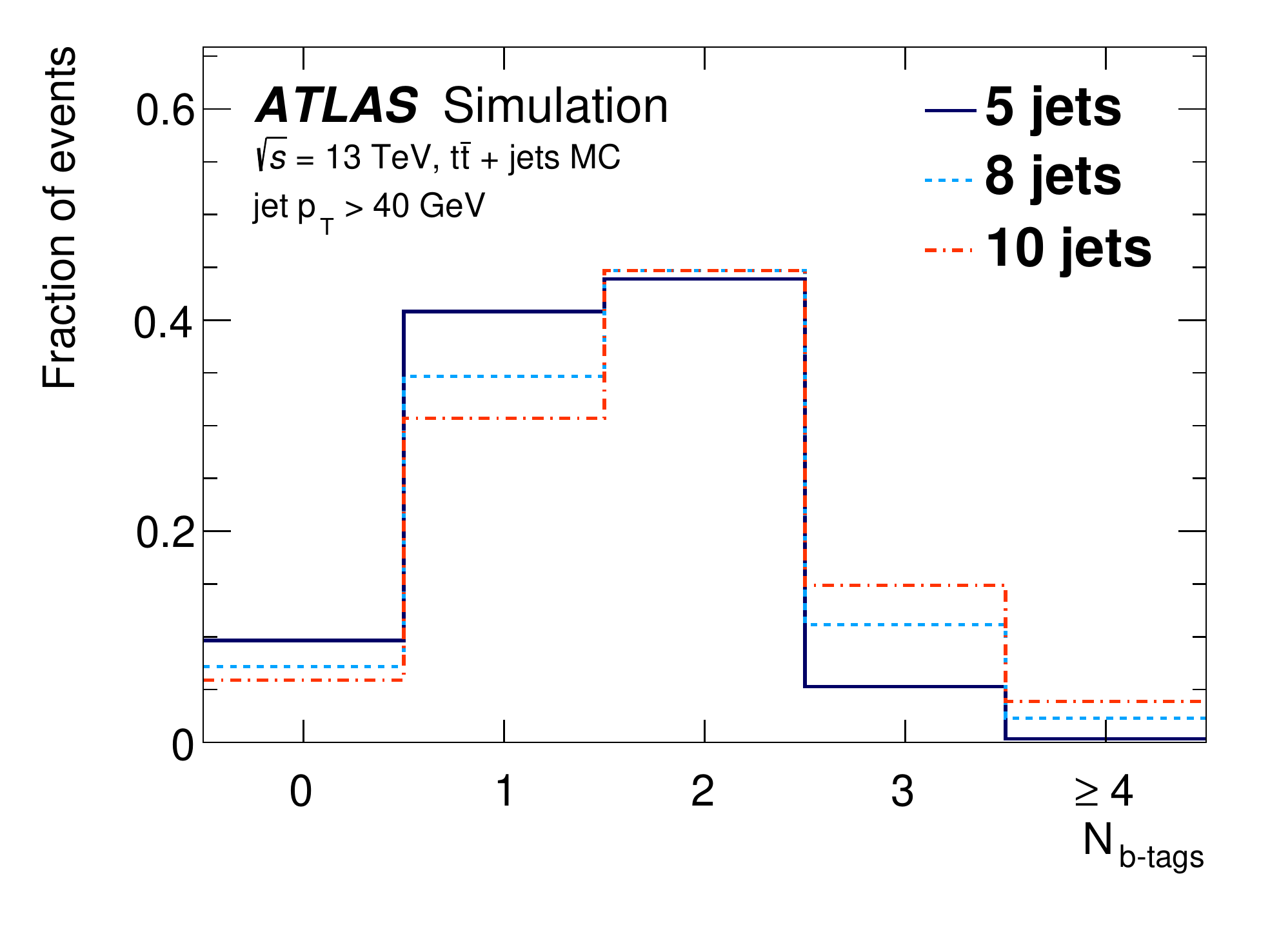}
\caption{
The normalized $b$-tag multiplicity distribution from $\ttbar$+jets MC simulation events with five, eight and ten jets (with \pT\ $> 40$~\GeV ).
}
\label{fig:BTagIllustration}
\end{figure}

A data-driven model is used to estimate the number of events from $\ttbar$+jets production in a given jet and $b$-tag multiplicity bin. The basic concept of this model is based on the extraction of an 
initial template of the $b$-tag multiplicity distribution in events with five jets and the parameterization of the evolution of this template to higher jet multiplicities. The absolute normalization for 
each jet slice is constrained in the fit as discussed later in this section.  
Figure~\ref{fig:BTagIllustration} shows the $b$-tag multiplicity distributions in $\ttbar$+jets MC simulation, for five-, eight- and ten-jet events, demonstrating how the distributions evolve as the number of jets increases. The background estimation parameterizes this effect and extracts the parameters describing the  evolution from a fit to the data. 

The extrapolation of the $b$-tag multiplicity distribution to higher jet multiplicities starts from the assumption that the difference between the $b$-tag multiplicity distribution in events with $j$ and $j+1$ jets arises mainly from the production of additional jets, and can be described by a fixed probability that the 
additional jet is $b$-tagged. Given the small mis-tag rate, this probability is dominated by the probability that the additional jet is a heavy-flavour jet which is $b$-tagged. In order to account for 
acceptance effects due to the different kinematics in events with high jet multiplicity, the probability of further $b$-tagged jets entering the acceptance is also taken into account. 
The extrapolation to one additional jet can be parameterized as:    
\begin{equation}
\begin{split}
  N^{\text{$\ttbar$+jets}}_{j, b} & = N^{\text{$\ttbar$+jets}}_j \cdot f_{j,b}, \\
  f_{(j+1),b} & = f_{j,b} \cdot x_0 + f_{j, (b-1)} \cdot x_1 + f_{j, (b-2)} \cdot x_2, 
\end{split}
\label{eqn:evolution}
\end{equation}
where $N^{\text{$\ttbar$+jets}}_j$ is the number of $\ttbar$+jets events with $j$ jets and $f_{j,b}$ is the fraction of $\ttbar$ events with $j$ jets of which $b$ are $b$-tagged. 
The parameters $x_i$ describe the probability of one additional jet to be either not $b$-tagged ($x_0$), $b$-tagged ($x_1$), or $b$-tagged and causing a second $b$-tagged jet to move into the 
fiducial acceptance ($x_2$). 
The latter is dominated by cases where the extra jet is a $b$-jet, influencing the event kinematics such that an additional $b$-jet, below the jet \pT\ threshold, enters the acceptance. 
Given that the $x_i$ parameters describe probabilities, the sum $\sum_{i} x_i$ is normalized to unity. Subsequent application of this parameterization produces 
a $b$-tag template for arbitrarily high jet multiplicities.

Studies based on MC simulated events with sample sizes corresponding to very large equivalent luminosities, as well as studies using fully efficient generator-level $b$-tagging, indicate the necessity to add a fit parameter that 
allows for correlated production of two $b$-tagged jets as may be expected with $b$-jet production from  gluon splitting. This is implemented by changing the evolution described in 
Eq.~(\ref{eqn:evolution}) such that any term with $x_1 \cdot x_1$ is replaced by $x_1 \cdot x_1 \cdot \rho_{11}$, where $\rho_{11}$ describes the correlated production of two $b$-tagged jets.

The initial $b$-tag multiplicity template is extracted from data events with five jets after subtracting all non-$\ttbar$ background processes, and is denoted by $f_{5,b}$ and scaled by the absolute 
normalization $N^{\text{$\ttbar$+jets}}_{5}$ in order to obtain the model in the five-jet bin:
\begin{equation*}
N^{\text{$\ttbar$+jets}}_{5,b}  = N^{\text{$\ttbar$+jets}}_{5} \cdot f_{5,b}, 
\end{equation*}
where the sum of $f_{5,b}$ over the five $b$-tag bins is normalized to unity.

The model described above is based on the assumption that any change of the $b$-tag multiplicity distribution is due to additional jet radiation with a  certain probability to lead to $b$-tagged jets. 
There is, however, also a small increase in the acceptance for $b$-jets produced in the decay of the \ttbar\ system, when increasing the jet multiplicity, due to the higher jet momentum on average. 
The effect amounts to up to 5\% in the one- and two-$b$-tag bins for high jet multiplicities, and is taken into account using a correction to the initial template extracted from simulated $\ttbar$ events.

As is the case for the $W/Z$+jets background, the normalization of the $\ttbar$ background in each jet slice is constrained using a scaling behaviour similar to that in Eq.~(\ref{eqn:scaling}). 
The parameterization is slightly modified to:
\begin{equation*}
N^{\text{$\ttbar$+jets}}_{j+1} / N^{\text{$\ttbar$+jets}}_j \equiv r^{\text{$\ttbar$+jets}}(j) = c^{\text{$\ttbar$+jets}}_0 +  c^{\text{$\ttbar$+jets}}_1 / (j + c^{\text{$\ttbar$+jets}}_2), 
\label{eqn:scalingttbar}
\end{equation*}
where the three parameters $c^{\text{$\ttbar$+jets}}_0$, $c^{\text{$\ttbar$+jets}}_1$ and $c^{\text{$\ttbar$+jets}}_2$ are extracted from a fit to the data. 
In this case, since $j$ is the total number of jets in the event, and not the number of jets produced in addition to the \ttbar\ system, the denominator $(j+1)$ in Eq.~(\ref{eqn:scaling}) is replaced by $(j+c^{\text{$\ttbar$+jets}}_2)$ to take into account the ambiguity in the counting of additional jets due to acceptance effects for the \ttbar\ decay products.       

\begin{figure}[htb]
\centering
\includegraphics[width=.7\textwidth]{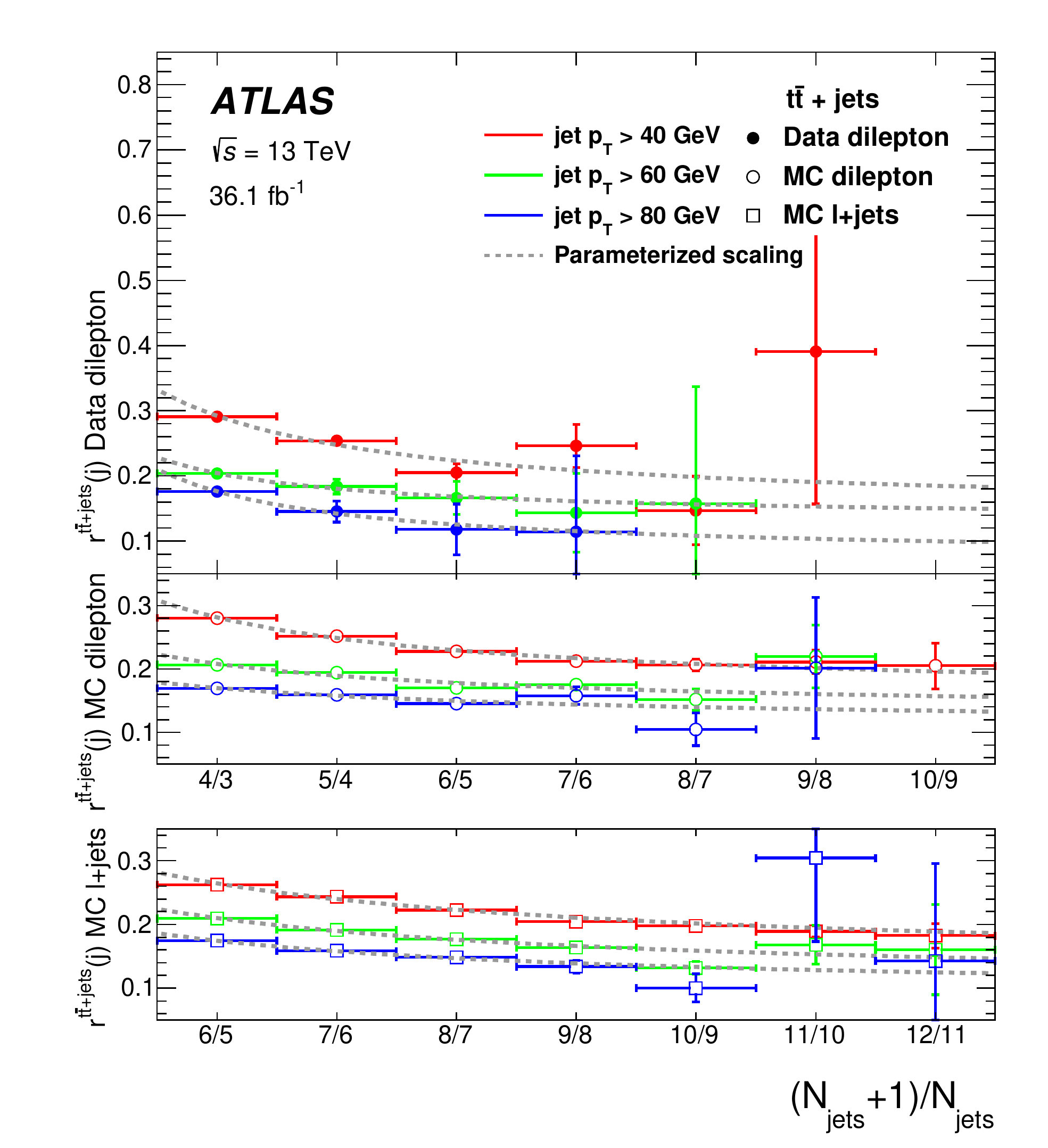}
\caption{
The ratio of the number of events with $(j+1)$ jets to the number with $j$ jets in dileptonic and semileptonic $\ttbar$+jets events, used to validate the jet-scaling parameterization. 
Each panel shows the ratio for data or MC simulation with the fitted parameterization overlaid as a line. In the case of pure ``staircase scaling'', the shown ratio would be a constant. 
The uncertainties shown are statistical. 
}
\label{fig:ttbar+jets}
\end{figure}

The scaling behaviour is tested in $\ttbar$+jets MC simulation (both with the nominal 
sample and the alternative sample described in Table~\ref{tab:MC}), 
and also in data with a dileptonic $\ttbar$+jets control sample. This sample is selected by requiring an electron candidate and a muon 
candidate in the event, with at least three jets of which at least one is $b$-tagged, and the small background predicted by MC simulation is subtracted. 
In this control region, the scaling behaviour can be tested for up to eight jets, but this corresponds to ten jets for a semileptonic $\ttbar$+jets sample (which is the dominant component of the $\ttbar$+jets background). Figure~\ref{fig:ttbar+jets} presents a comparison of the scaling behaviour in data and MC simulation compared to a fit of the parameterization used and shows that the assumed function describes
 the data and MC simulation well for the jet-multiplicity range relevant to this search. 

As for the $W/Z$+jets background estimate, the $\ttbar$+jets background model is used to predict the yield in the highest jet-multiplicity bin by iterating to higher jet multiplicities and summing these
 contributions to give the inclusive yield.

The zero-$b$-tag component of the initial $\ttbar$ template, which is extracted from events with five jets, exhibits an anti-correlation with the absolute $W$+jets normalization, which is extracted in the 
same bin. The control regions separated in leading-lepton charge, detailed in Section~\ref{sec:bkd_vjets}, provide a handle to extract the absolute $W$+jets normalization. 
The remaining anti-correlation does not affect the total background estimate. 
For these control regions, the $\ttbar$+jets process is assumed to be charge symmetric and the model is simply split into two halves for these bins. 

\subsection{Multi-jet events}
\label{subsec:bkd_qcd}
The contribution from multi-jet production with a fake or non-prompt (FNP) lepton (such as hadrons misidentified as leptons, leptons originating from the decay of heavy-flavour hadrons, and electrons from photon 
conversions), constitutes a minor but non-negligible background, especially in the lower jet slices. It is estimated from the data with a matrix method similar to that described in 
Ref.~\cite{paperSS3L}. In this method, two types of lepton identification criteria are defined: ``tight'', corresponding to the default lepton criteria described in Section~\ref{sec:objects}, and 
``loose'', corresponding to baseline leptons after overlap removal. The matrix method relates the number of events containing prompt or FNP leptons to the number of observed events with tight or 
loose-not-tight leptons using the probability for loose-prompt or loose-FNP leptons to satisfy the tight criteria. The probability for loose-prompt leptons to satisfy the tight selection criteria is 
obtained using a $Z\to\ell\ell$ data sample and is modelled as a function of the lepton $\pt$.  The probability for loose FNP leptons to satisfy the tight selection criteria is determined from a data 
control region enriched in non-prompt leptons requiring a loose lepton, multiple jets, low \met~\cite{ATL-PHYS-PUB-2015-027,ATL-PHYS-PUB-2015-023} and low transverse mass.\footnote{The transverse mass of the lepton--\met system is defined as: $\mt^2=2 \pT^{\ell}\met(1 - \cos(\Delta\phi(\ell,\met))$).}
 The efficiencies are measured as a 
function of lepton candidate \pt after subtracting the contribution from prompt-lepton processes and are assumed to be independent of the jet multiplicity.\footnote{To minimize the dependence on the number of jets, 
the event selection considers only the leading-\pT\ baseline lepton when checking the more stringent identification and isolation criteria of the ``tight'' lepton definitions.}

\subsection{Small backgrounds}
\label{subsec:bkd_mc}
The small background contributions from diboson production, single-top production, $\ttbar$ production in association with a vector/Higgs boson (labeled $\ttbar V/H$) and SM four-top-quark production 
are estimated using MC simulation. In all but the highest jet slices considered, the sum of these backgrounds contributes not more than 10\% of the SM expectation in any of the 
$b$-tag bins; for the highest jet slices this can rise up to 35\% .

\section{Fit configuration and validation}
\label{subsec:bkd_validation}
For each jet \pT\ threshold, the search results are determined from a simultaneous likelihood fit. The likelihood is built as the product of Poisson probability terms describing the observed 
numbers of events in the different bins and Gaussian distributions constraining the nuisance parameters associated with the systematic uncertainties. The widths of the Gaussian distributions correspond to the sizes of these 
uncertainties. Poisson distributions are used to constrain the nuisance parameters for MC simulation and data control region statistical uncertainties. Correlations of a given nuisance parameter between the different background sources 
and the signal are taken into account when relevant. The systematic uncertainties are not constrained by the data in the fit procedure.

The likelihood is configured differently for the model-dependent and model-independent hypothesis tests.  The former is used to derive exclusion limits for a specific BSM model, and 
the full set of bins (for example $5$ to $12$-inclusive jet multiplicity bins, and $0$ to $4$-inclusive $b$-jet bins for the 40 \GeV\ jet \pT threshold) is employed in the likelihood. The signal 
contribution, as predicted by the given BSM model, is considered in all bins and is scaled by one common signal-strength parameter.
The number of freely floating parameters in the background model is 15.  There are four parameters in the $W/Z$+jets model: the two jet-scaling parameters ($c_0$, $c_1$), and the normalizations of the $W$+jets and $Z$+jets 
events in the five-jet region ($N^{\text{$W$+jets}}_{5}$, $N^{\text{$Z$+jets}}_{5}$). In addition, there are 11 parameters in the $\ttbar$+jets background model: one for the normalization in the five-jet slice 
($N^{\text{$\ttbar$+jets}}_5$), three for the normalization scaling ($c^{\text{$\ttbar$+jets}}_0$, $c^{\text{$\ttbar$+jets}}_1$, $c^{\text{$\ttbar$+jets}}_2$), four for the initial $b$-tag multiplicity 
template ($f_{5,b}$, $b=1$--$4$), and three for the evolution parameters ($x_1$, $x_2$ and $\rho_{11}$), taking into account the constraints: $x_0=1-x_1-x_2$, and $f_{5,0} = 1 - \sum_{b=1}^{\ge 4} f_{5,b}$. The number 
of fitted bins\footnote{For example, for the 60 and 80~\GeV\ jet \pT\ thresholds, there are five $b$-tag multiplicity bins in the eight-to-ten-jet slices, and seven bins (the zero-$b$-tag bin is split into three bins 
for each of the $W/Z$ control regions) in the five-, six- and seven-jet slices, giving 36 bins in total.} varies between 36 and 46 depending on the highest jet-multiplicity bin used, leading to an over-constrained 
system in all cases. 

The model-independent test is used to search for, and to set generic exclusion limits on, the potential contribution from a hypothetical BSM signal in the phase-space region probed by this analysis. For this purpose, 
dedicated signal regions are defined which could be populated by such a signal, and where the SM contribution is expected to be small. The SR selections are defined as requiring exactly zero or 
at least three $b$-tags (labelled 0b, or 3b respectively) for a given minimum number of jets J, and for a jet \pT threshold X, with each SR labelled as X-0b-J or X-3b-J. For each jet \pT\ threshold, six SRs 
are defined as follows: 
\begin{itemize}
\item For the 40~\GeV\ jet \pT threshold: 40-0b-10, 40-3b-10, 40-0b-11, 40-3b-11, 40-0b-12, 40-3b-12.  
\item For the 60~\GeV\ jet \pT threshold: 60-0b-8, 60-3b-8, 60-0b-9, 60-3b-9, 60-0b-10, 60-3b-10.  
\item For the 80~\GeV\ jet \pT threshold: 80-0b-8, 80-3b-8, 80-0b-9, 80-3b-9, 80-0b-10, 80-3b-10. 
\end{itemize}
The SRs therefore overlap and an event can enter more than one SR. Due to the efficiency of the $b$-tagging algorithm used, signal models with large $b$-tag multiplicities can have significant 
contamination in the two-$b$-tag bins, which can bias the $\ttbar$+jets background estimate and reduce the sensitivity of the search. To reduce this effect, for the SRs with $\ge 3$ $b$-tags, the two-$b$-tag bin 
is not included in the fit for the highest jet slice in each SR.\footnote{For example, for the 60-0b-8 and 80-0b-8 SRs all bins with five, six or seven jets are included in the fit, as well as the one-, two-, three- and four-or-more-$b$-tag bins with at least eight jets. Whereas for the 60-3b-8 and 80-3b-8 SRs all bins with five, six or seven jets are included in the fit, as well as the zero- and one-$b$-tag bins with at least eight jets.}

For the model-independent hypothesis tests, a separate likelihood fit is performed for each SR. A potential signal contribution is considered in the given SR bin only.  The number of freely floating 
parameters in the background model is 15, whereas the number of observables varies between 23 (for SRs 60-3b-8 and 80-3b-8) and 45 (for SR 40-0b-12), so the system is also always over-constrained. 

The fit set-up was extensively tested using MC simulated events, and was demonstrated to give a negligible bias in the fitted yields, both in the case where the background-only distributions 
are fit, or when a signal is injected into the fitted data. These tests were carried out with the nominal MC samples as well as the alternative samples described in Table~\ref{tab:MC}. 
In addition, when fitting the data the fitted parameter values and their inter-correlations were studied in detail and found to be in agreement with the expectation based on MC simulation. 
The jet-reconstruction stability at high multiplicities was validated by comparing jets with track-jets that are clustered from ID tracks with a radius parameter of 0.2. The ratio of the 
multiplicities of track-jets and jets, which is sensitive to jet-merging effects, was found to be stable up to the highest jet multiplicities studied. 
The estimate of the multi-jet background was validated in data regions enriched in FNP leptons, and was found to describe the data within the quoted uncertainties.

%% file: systematics.tex
\section{Systematic uncertainties}
\label{sec:syst}

The dominant backgrounds are estimated from the data without the use of MC simulation, and therefore the main systematic uncertainties related to the estimation of these backgrounds arise from the 
assumptions made in the $W/Z$+jets, $\ttbar$+jets and multi-jet background estimates. Uncertainties related to the theoretical modelling of the specific processes and due to the modelling of the 
detector response in simulated events are only relevant for the minor backgrounds,
which are taken from MC simulation, and for the estimates of the signal yields after selections.

For the $W/Z$+jets background estimation, the uncertainty related to the assumed scaling behaviour is taken from studies of this behaviour in $W$+jets and $Z$+jets MC simulation, as well as in 
$\gamma$+jets and multi-jet data control regions chosen to be kinematically similar to the search selection (see Figure~\ref{fig:v+jets}). No evidence is seen for a deviation from the assumed scaling 
behaviour and the statistical precision of these methods is used as an uncertainty (up to 18\% for the highest jet-multiplicity bins). The expected uncertainty of the charge asymmetry for $W$+jets 
production is $3$--$5\%$ from PDF variations~\cite{Botje:2011sn}, but in the seven-jet region, the uncertainty is dominated by the limited number of MC events (up to 10\% for the 80~\GeV\ jet \pT\ threshold). The 
uncertainty in the shape of the $b$-tag multiplicity distribution in $W$+jets and $Z$+jets events is derived by comparing different MC generator set-ups (e.g. varying the renormalization and factorization
 scale and the parton-shower model parameters). It is seen to grow as a function of jet multiplicity and is about 50\% for events with five jets, after which the MC statistical uncertainty becomes very large. A conservative 
uncertainty of 100\% is therefore assigned to the fractional contribution from $W$+$b$ and $W$+$c$ events for all jet slices considered, which has a very small impact on the final result as the background from $W$ boson production with additional heavy flavour jets is small compared to that from top quark pair production. 
In addition, the uncertainties related to the $b$-tagging efficiency and mis-tag rate are taken into account in the uncertainty in the $W/Z$+jets $b$-tag template. 

The uncertainties related to the $\ttbar$+jets background estimation primarily relate to the number of events in the data regions used for the fit. As mentioned in Section~\ref{subsec:bkd_tt}, the 
method shows good closure using simulated events, so no systematic uncertainty related to these studies is assigned. There is a small uncertainty related to the acceptance correction for the 
initial $b$-tag multiplicity template, which is derived by varying the MC generator set-up for the $t\bar{t}$ sample used to estimate the correction. This leads to a 3\% uncertainty in the correction 
and has no significant effect on the final uncertainty. 
The uncertainty related to the parameterization of the scaling of the $\ttbar$+jets background with jet multiplicity is determined with MC simulation closure tests. 
The validation of the method presented in Figure~\ref{fig:ttbar+jets} shows that the parameterization describes the data and MC simulation well. The uncertainties assigned vary  
from 3\% (at 8 jets) to 33\% (at 12 jets) for the 40~\GeV\ jet \pT threshold, and from 10\% (at 8 jets) to 60\% (at 10 jets) for the 80~\GeV\ jet \pT threshold.
These are estimated by studying the closure of the method in different MC samples (including using alternative MC generators, and varying the event selection) and are of similar size to the statistical uncertainty from the data validation. 

The dominant uncertainties in the multi-jet background estimate arise from the number of data events in the control regions, uncertainties related to the subtraction of electroweak backgrounds from 
these control regions (here a 20\% uncertainty is applied to the expected yield of the backgrounds in the control regions) and uncertainties to cover the possible dependencies of the real- and fake-lepton 
efficiencies~\cite{paperSS3L} on variables other than lepton \pT (for example the dependence on the number of jets in the event). The total uncertainty in the multi-jet background yields is about 50\%.
   
The uncertainty in the expected yields of the minor backgrounds includes theoretical uncertainties in the cross-sections and in the modelling of the kinematics by the MC generator, as well as experimental
 uncertainties related to the modelling of the detector response in the simulation. The uncertainties assigned to cover the theoretical estimate of these backgrounds in the relevant regions are 50\%, 
100\% and 30\% for diboson, single top-quark, and $\ttbar V/H$ production, respectively. 

The final uncertainty in the background estimate in the SRs is dominated by the statistical uncertainty related to the number of data events in the different bins, and other systematic uncertainties do not contribute significantly. 

The uncertainties assigned to the expected signal yield for the SUSY benchmark processes considered include the experimental uncertainties related to the detector modelling, which are dominated by the 
modelling of the jet energy scale and the $b$-tagging efficiencies and mis-tagging rates. For example, for a signal model with four $b$-quarks the $b$-tagging uncertainties are $\approx$10\%, and the jet related uncertainties are typically $\approx$5\%. The uncertainty in the signal cross-sections used is discussed in 
Section~\ref{sec:signal}. The uncertainty in the signal yields related to the modelling of additional jet radiation is studied by varying the factorization, renormalization, and jet-matching scales as well as the parton-shower tune in the simulation. The corresponding uncertainty is small for most of the signal parameter space, but increases to up to 25\% for very light or very heavy LSPs where the contribution from additional jet radiation is relevant.

%% file: results.tex
\section{Results}
\label{sec:results}

Results are provided both as model-independent limits on the contribution from BSM physics to the dedicated signal regions and in the context of the four SUSY benchmark models discussed in 
Section~\ref{sec:signal}. As previously mentioned, different fit set-ups are used for these two sets of results. In all cases, the profile-likelihood-ratio test~\cite{Cowan:2010js} is used to establish 
95\% confidence intervals using the CL$_\mathrm{s}$ prescription~\cite{Read_CLs}. 

\begin{figure}[htbp]
\centering
\includegraphics[width=.32\textwidth]{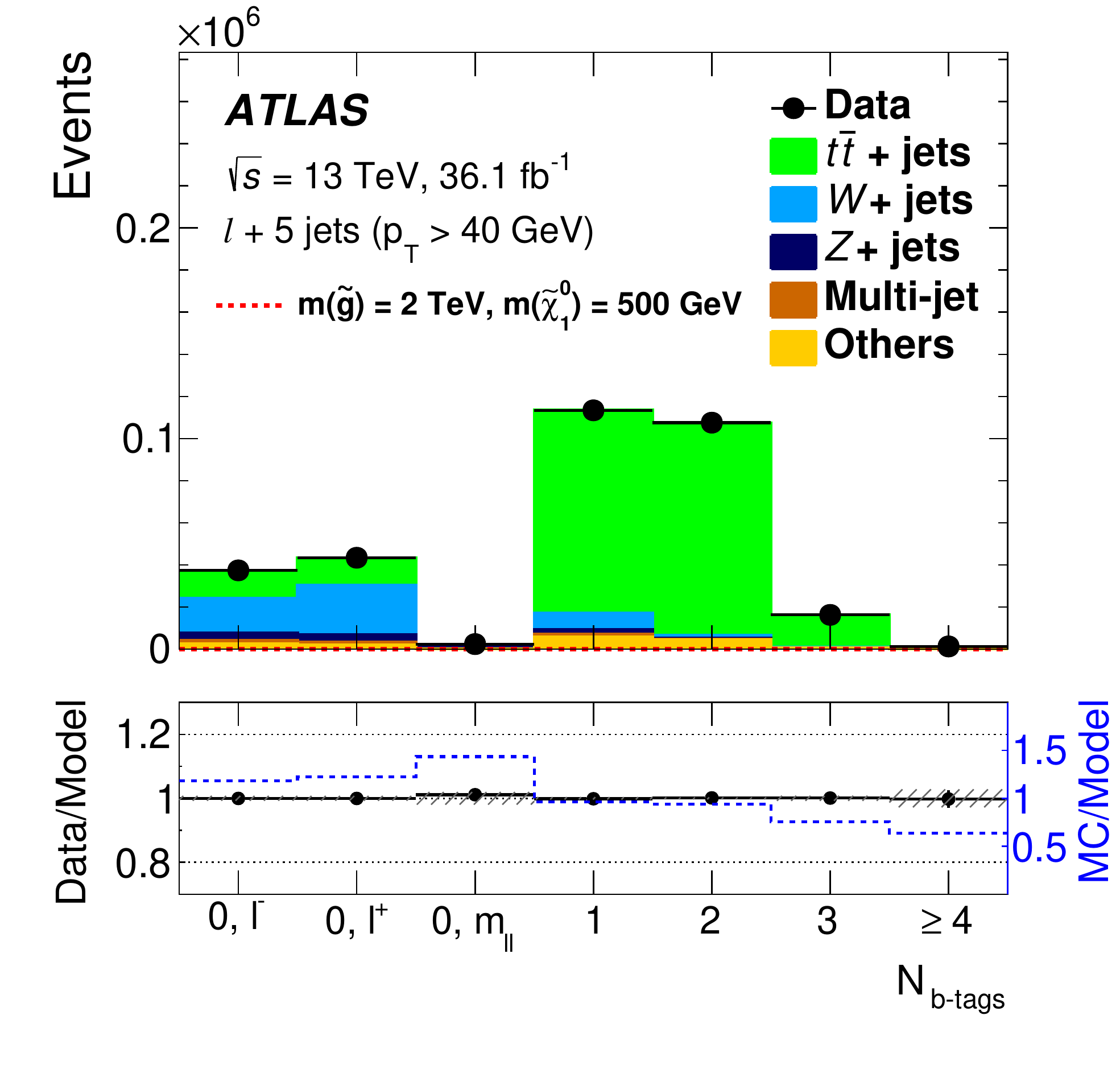}
\includegraphics[width=.32\textwidth]{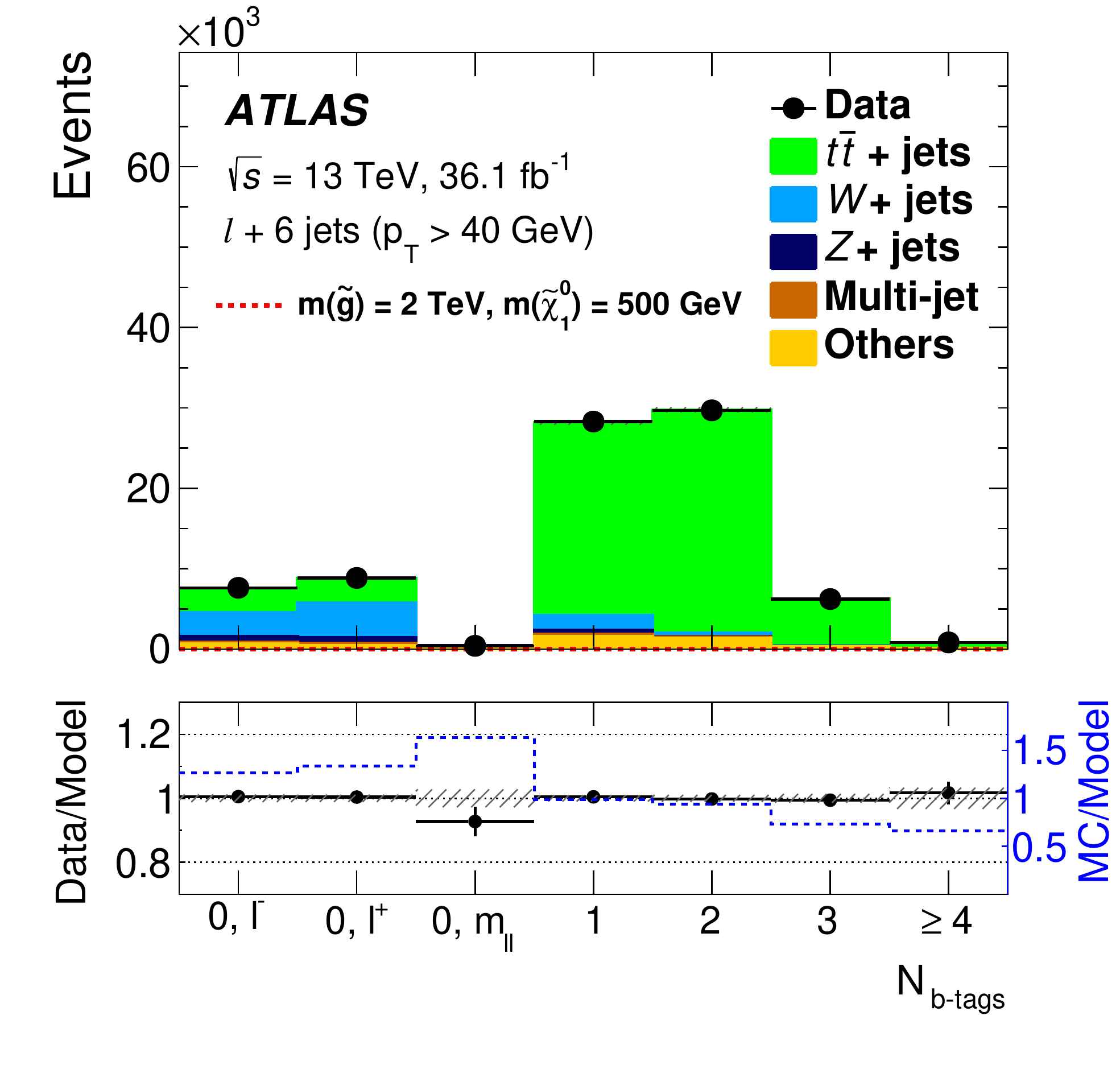}
\includegraphics[width=.32\textwidth]{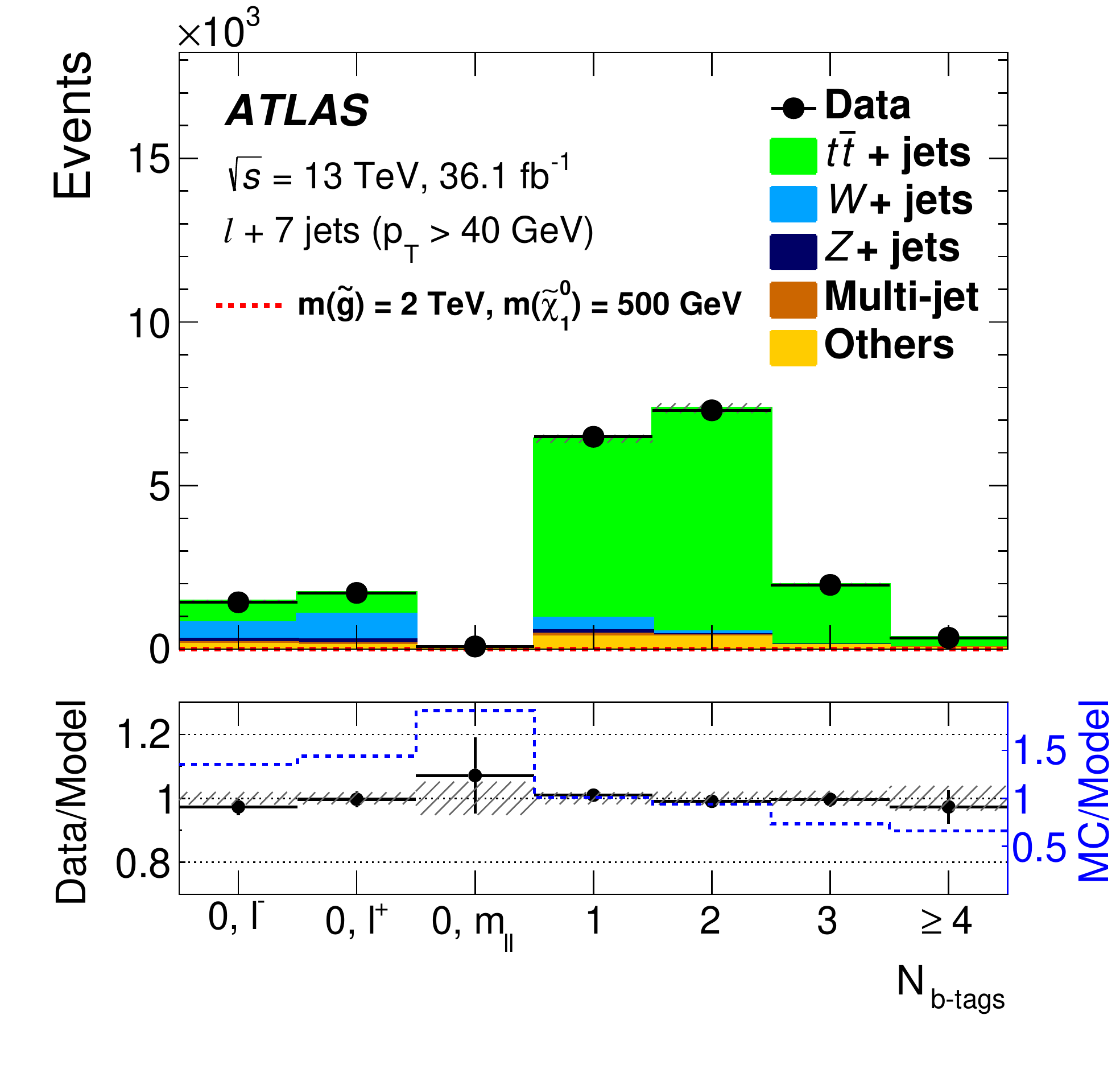}\\
\includegraphics[width=.32\textwidth]{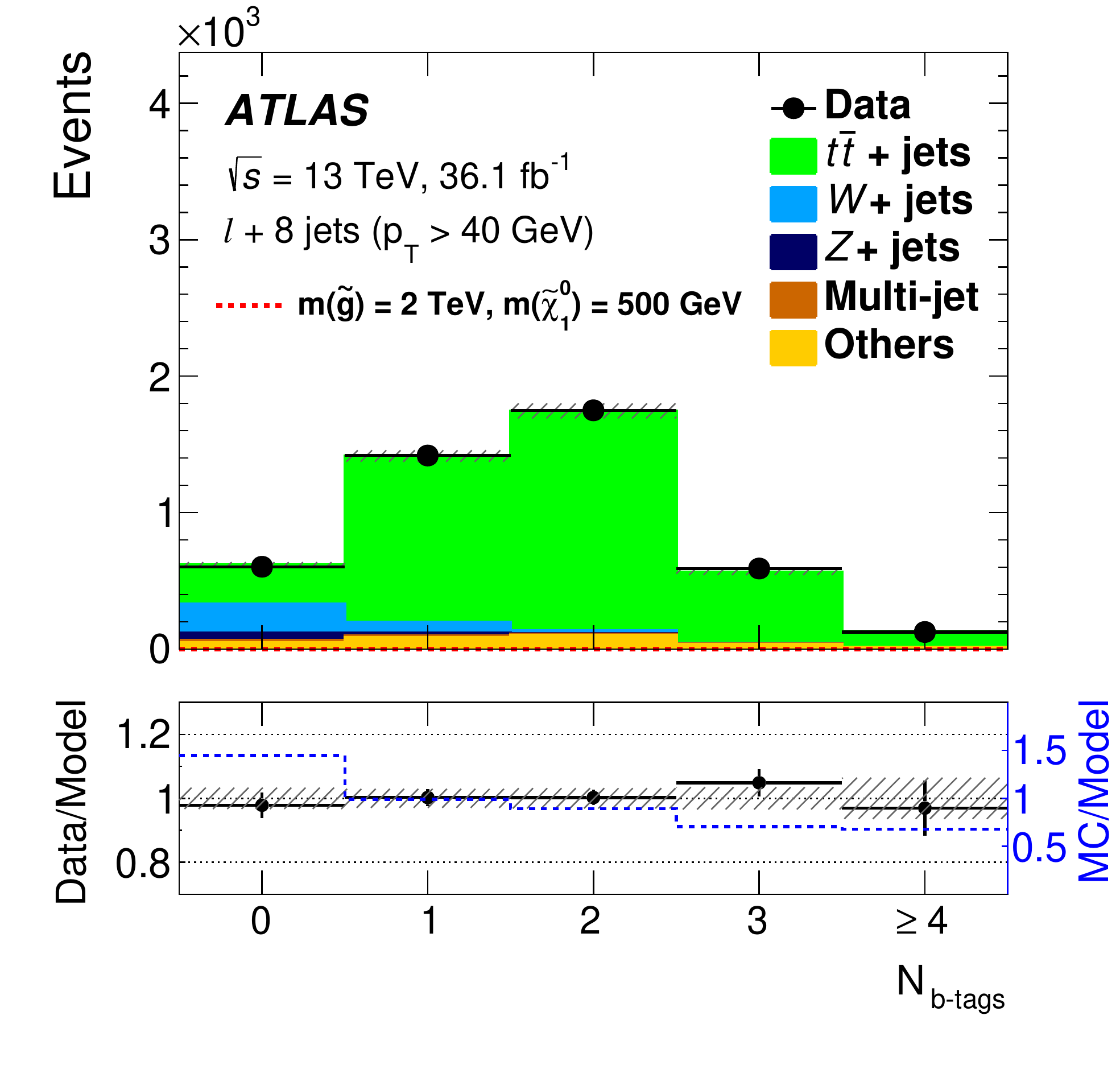}
\includegraphics[width=.32\textwidth]{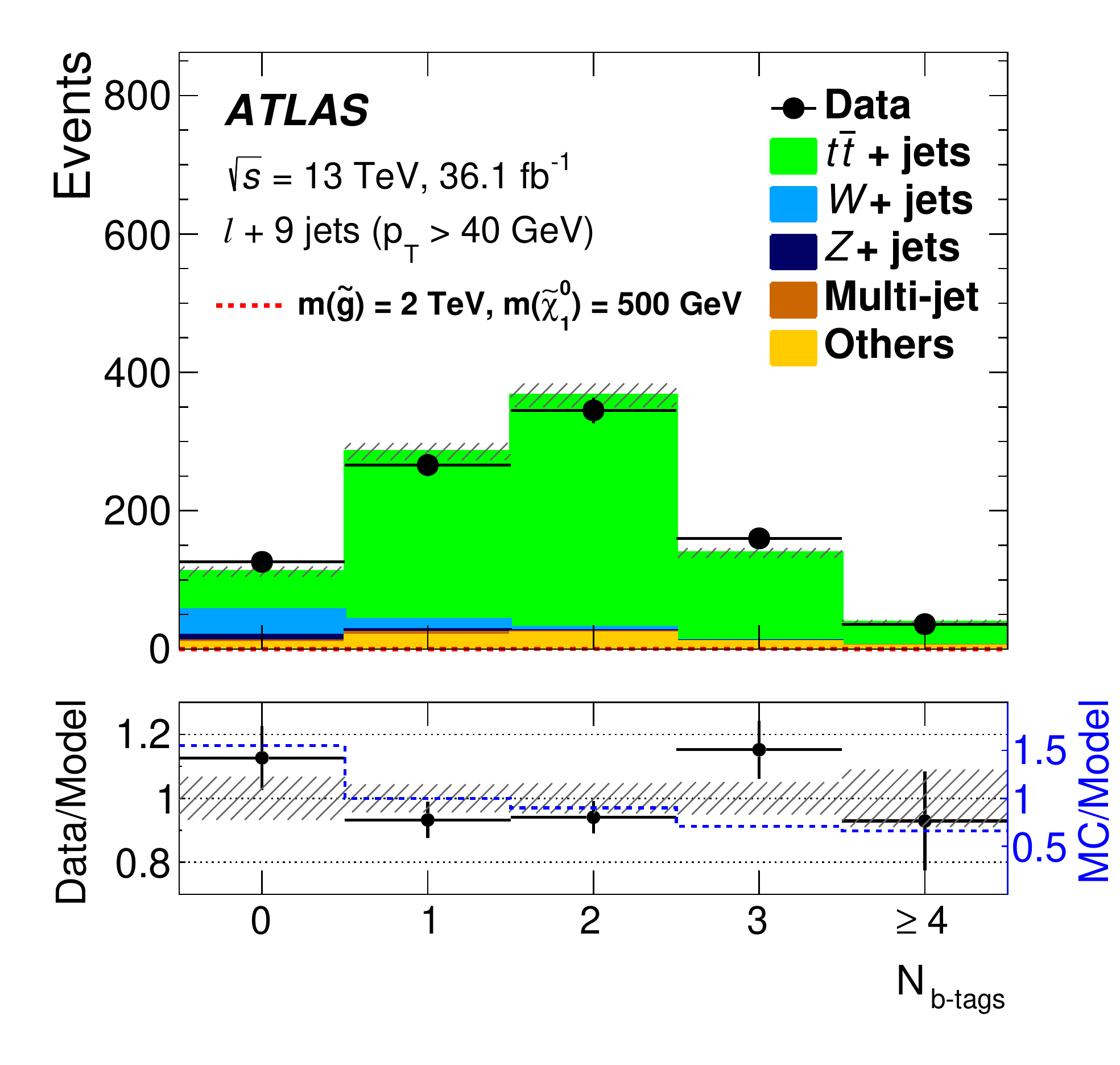}
\includegraphics[width=.32\textwidth]{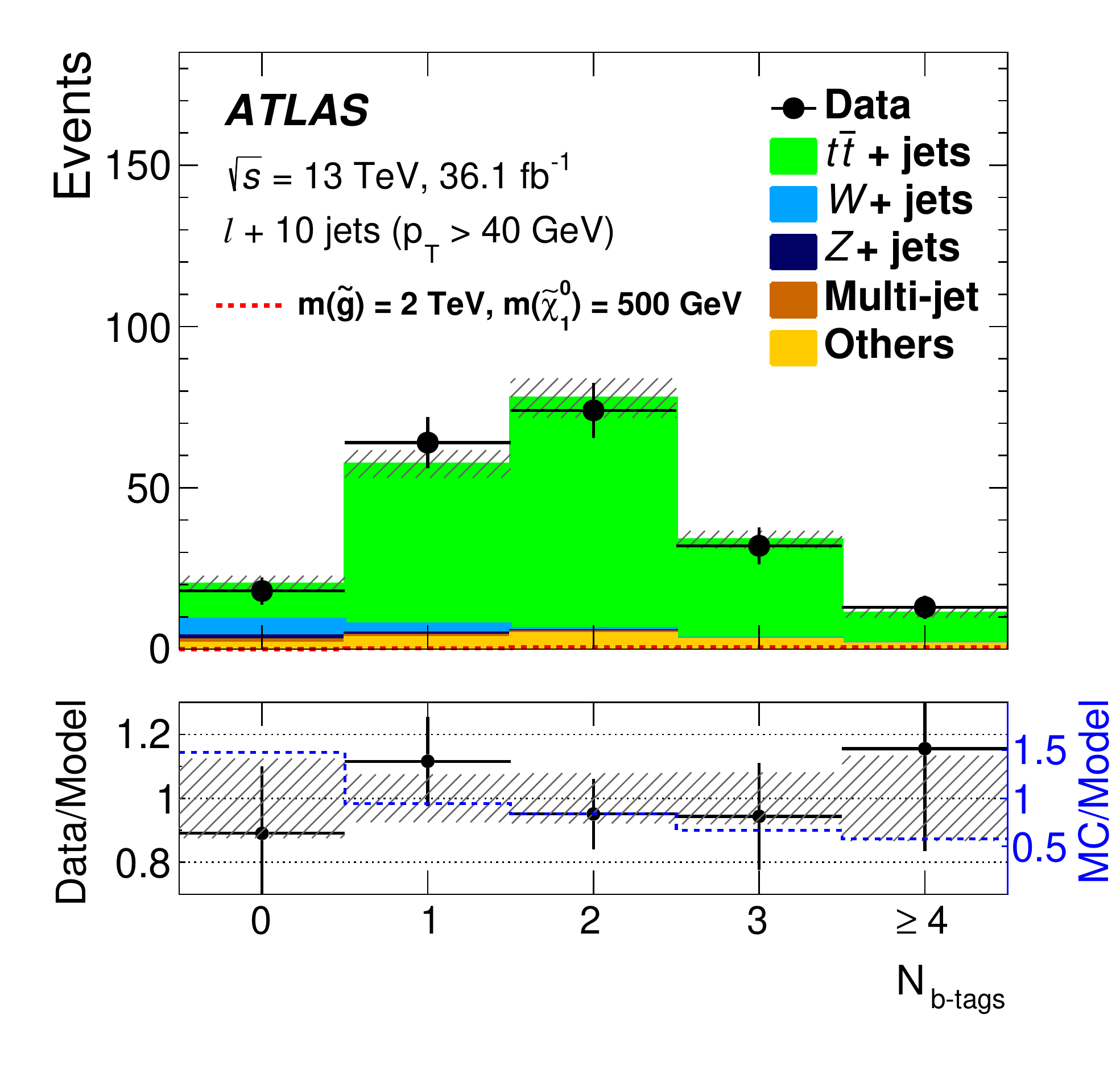}\\
\includegraphics[width=.32\textwidth]{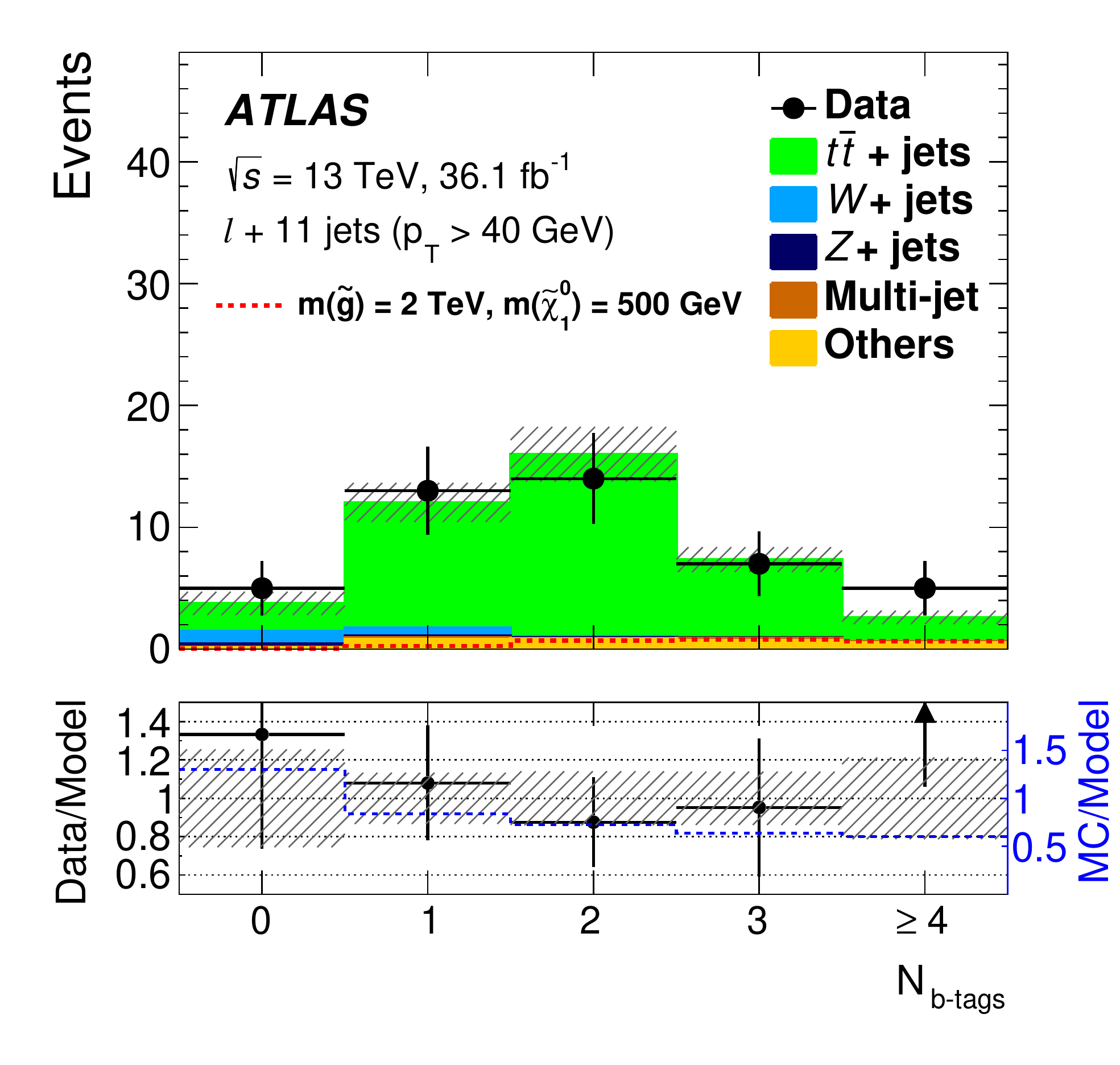}
\includegraphics[width=.32\textwidth]{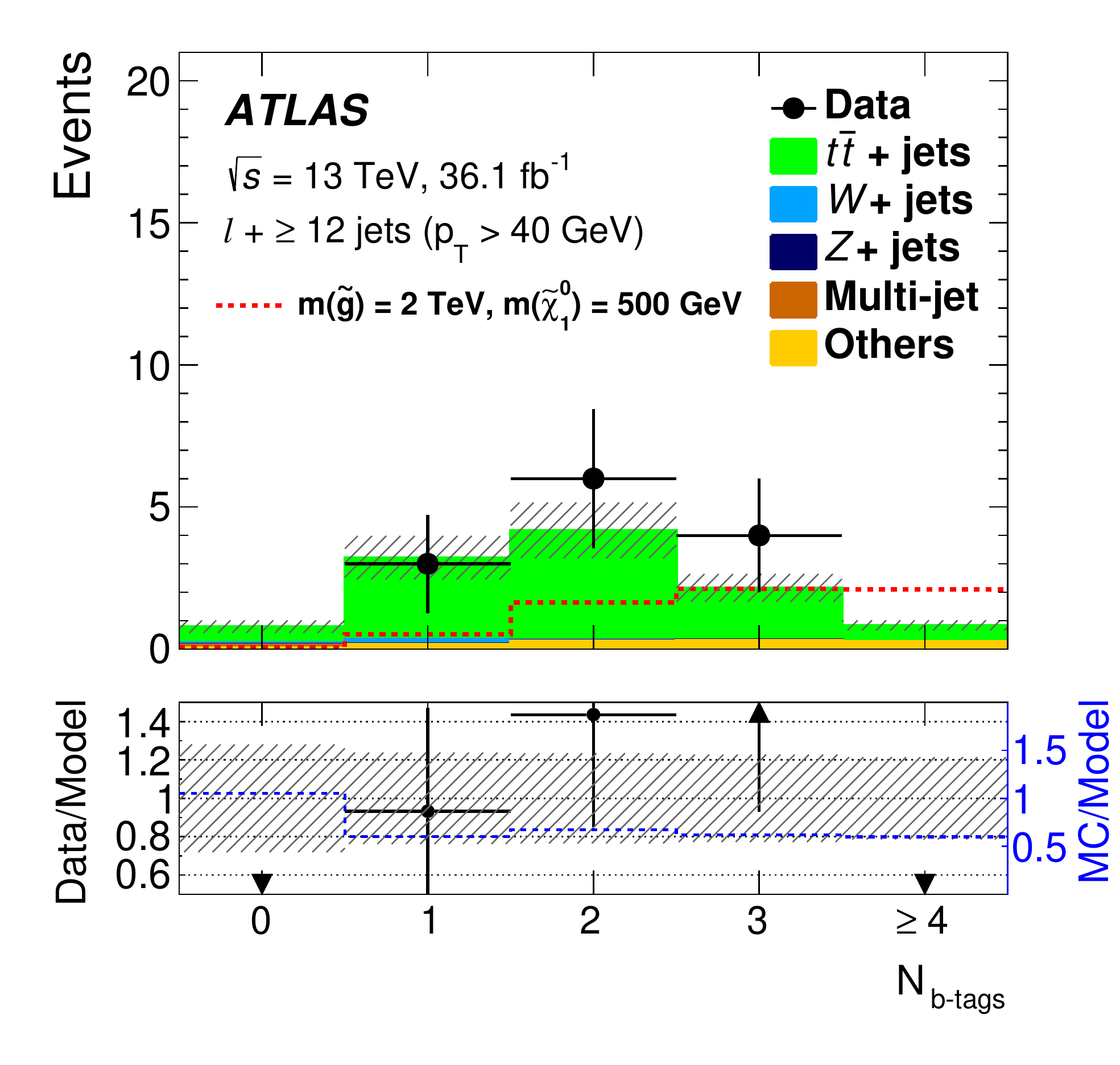}
\caption{The expected background and observed data in the different jet and $b$-tag multiplicity bins for the 40~\GeV\ jet \pT\ threshold. The background shown is estimated by including all bins in the 
fit. For the five-, six- and seven-jet slices the control regions used to estimate the $W$+jets and $Z$+jets normalizations are also shown (labelled $\ell^-$, $\ell^+$, and $m_{\ell \ell}$).  An example 
signal for the $\gluino\to t\bar t\ninoone \to  t\bar t u d s$ model with $m_{\gluino}$ = 2000~\GeV\ and $m_{\ninoone}$ = 500~\GeV\ is also overlaid (although its contribution is very small with this 
jet \pT\ threshold).
The bottom panels show the ratio of the observed data to the expected background, as well as the ratio of the prediction from MC simulation to the expected background.  
All uncertainties, which can be correlated across the bins, are included in the error bands (shaded regions).}
\label{fig:postfit_40_77_unblind}
\end{figure}

\begin{figure}[htbp]
\centering
\includegraphics[width=.43\textwidth]{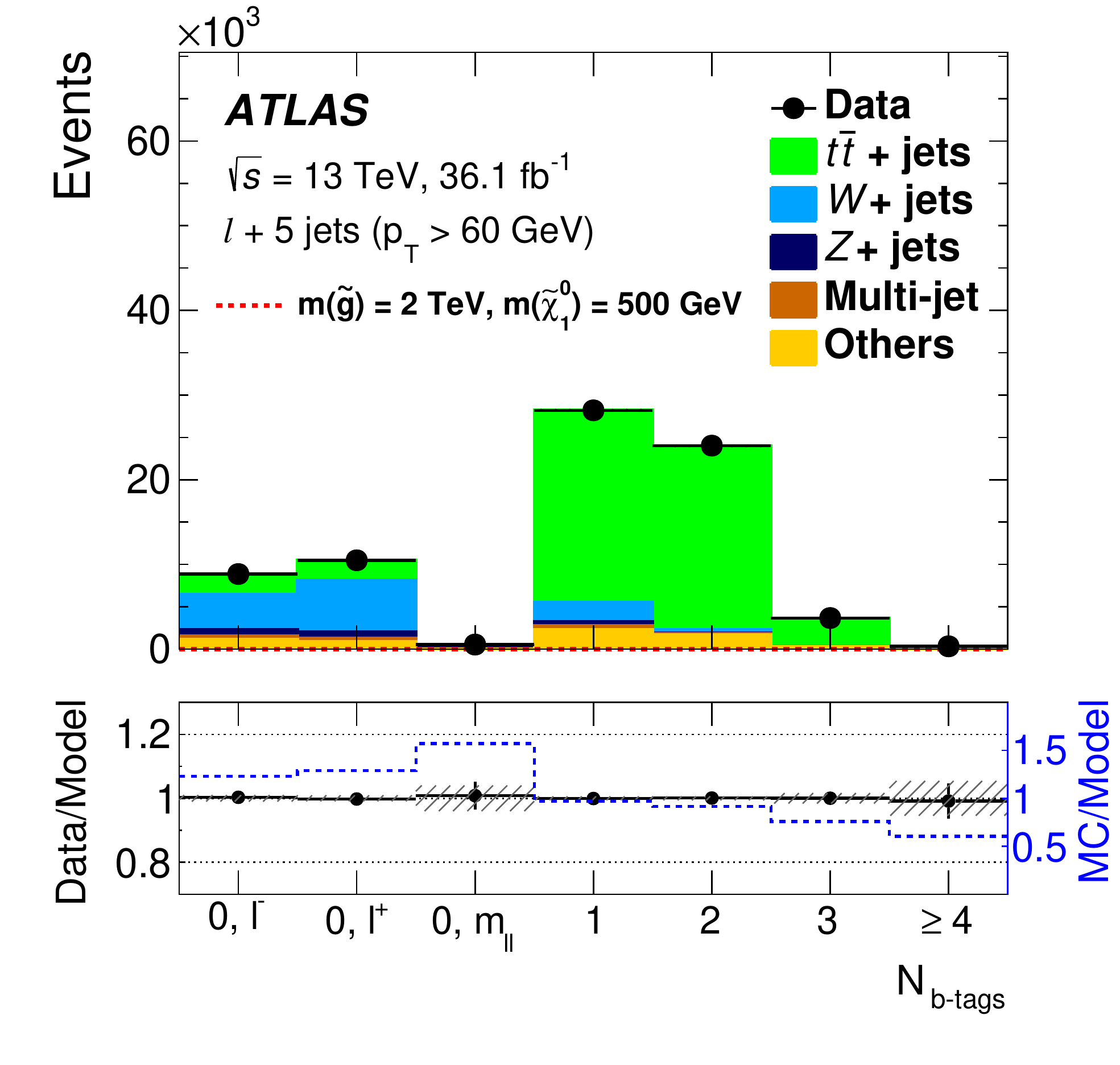}
\includegraphics[width=.43\textwidth]{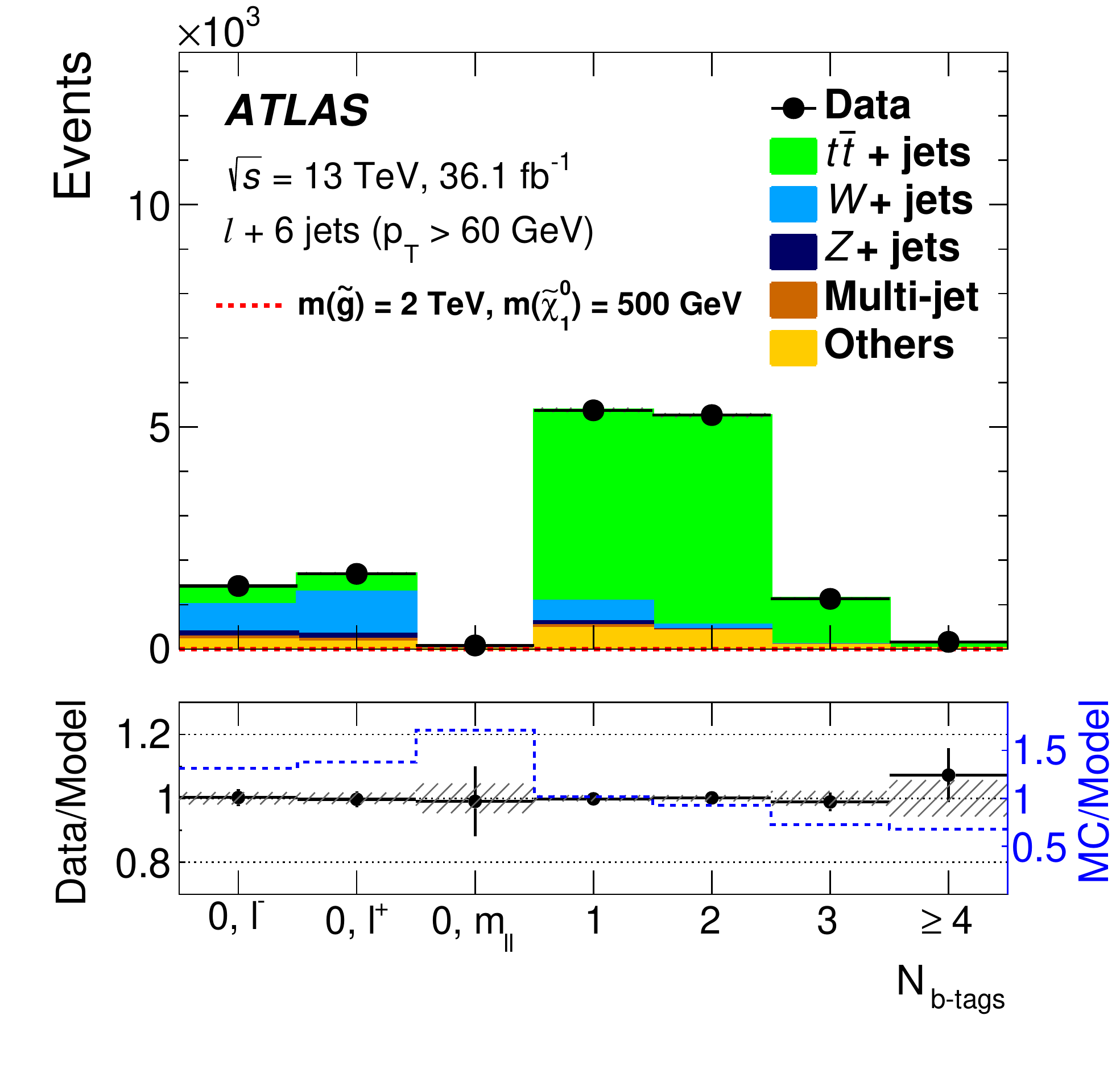}\\
\includegraphics[width=.43\textwidth]{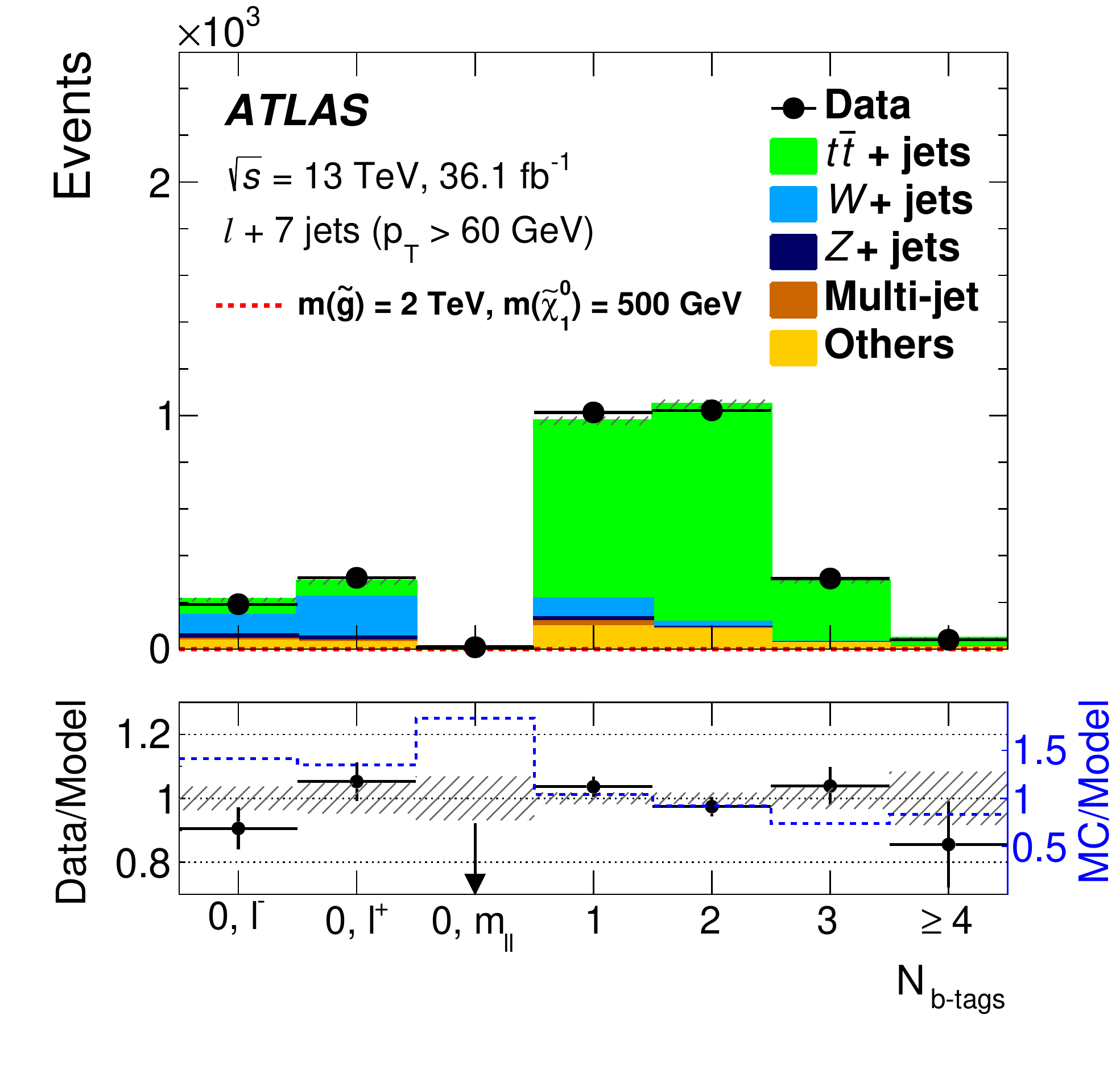}
\includegraphics[width=.43\textwidth]{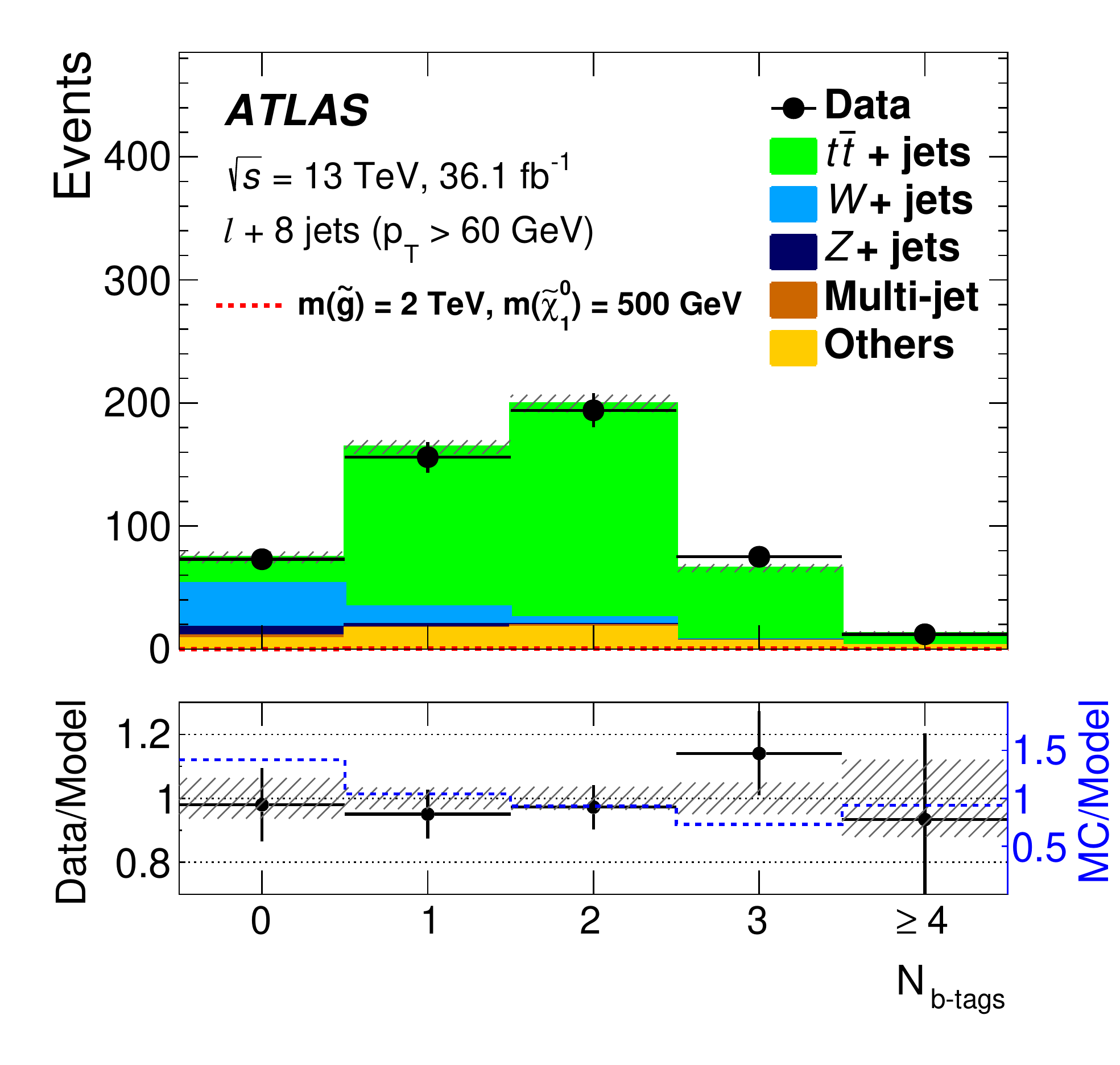}\\
\includegraphics[width=.43\textwidth]{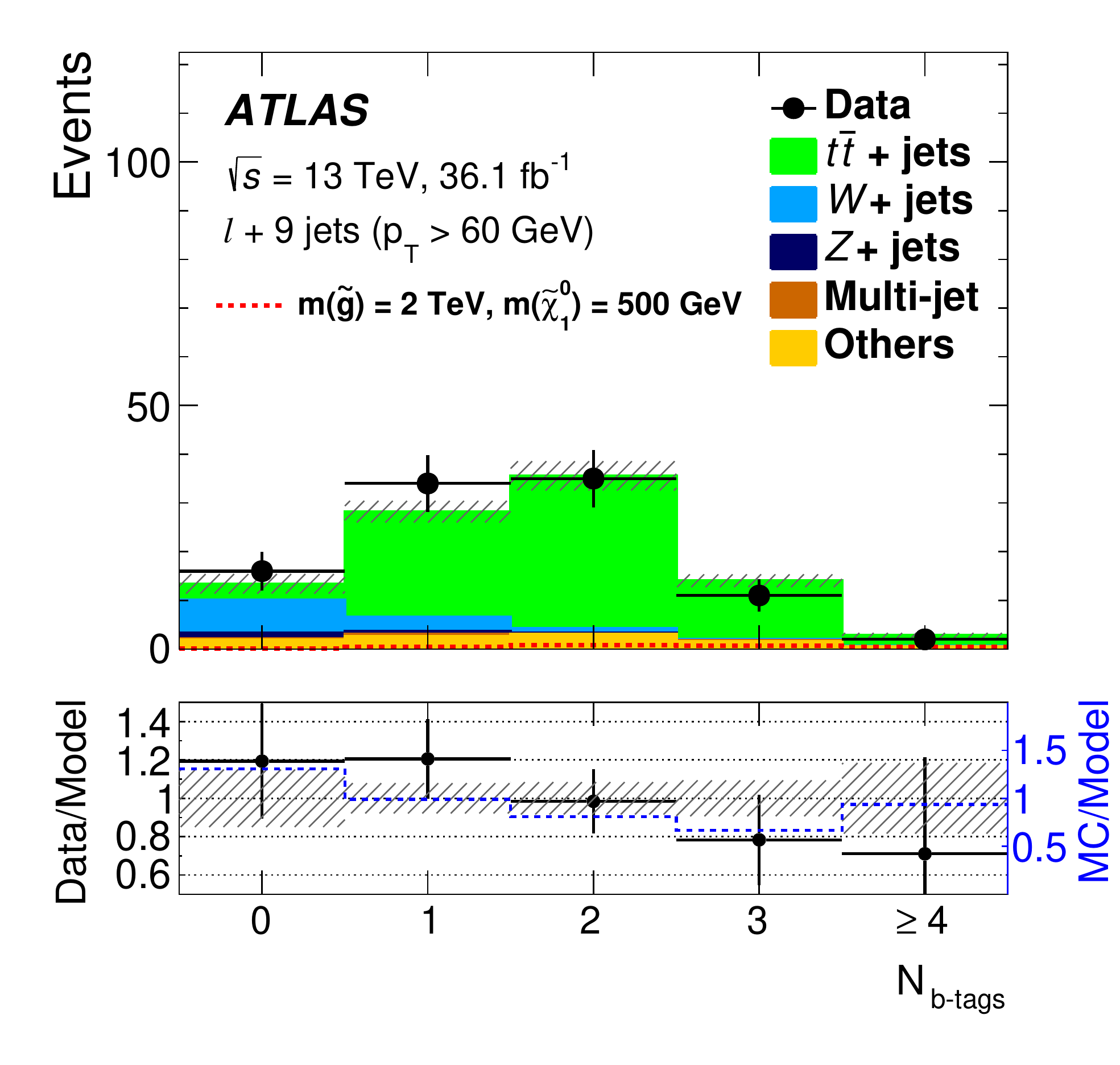}
\includegraphics[width=.43\textwidth]{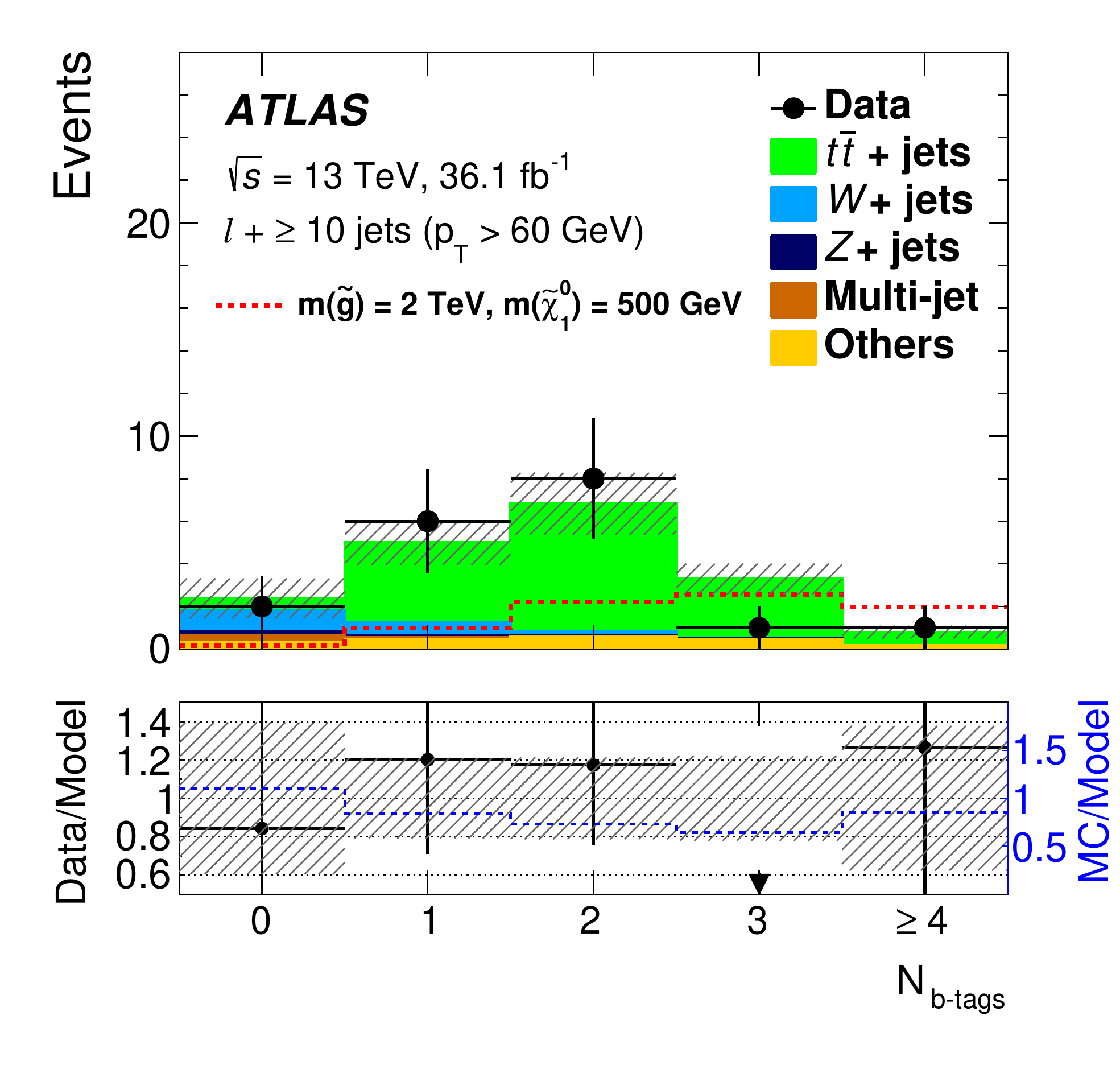}\\
\caption{The expected background and observed data in the different jet and $b$-tag multiplicity bins for the 60~\GeV\ jet \pT\ threshold. The background shown is estimated by including all bins in the 
fit. For the five-, six- and seven-jet slices the control regions used to estimate the $W$+jets and $Z$+jets normalizations are also shown (labelled $\ell^-$, $\ell^+$, and $m_{\ell \ell}$). An example 
signal for the $\gluino\to t\bar t\ninoone \to  t\bar t u d s$ model with $m_{\gluino}$ = 2000~\GeV\ and $m_{\ninoone}$ = 500~\GeV\ is also overlaid (although its contribution is very small in most of the 
jet multiplicity slices shown).
The bottom panels show the ratio of the observed data to the expected background, as well as the ratio of the prediction from MC simulation to the expected background. 
All uncertainties, which can be correlated across the bins, are included in the error bands (shaded regions).}

\label{fig:postfit_60_77_unblind}
\end{figure}

\begin{figure}[htbp]
\centering
\includegraphics[width=.43\textwidth]{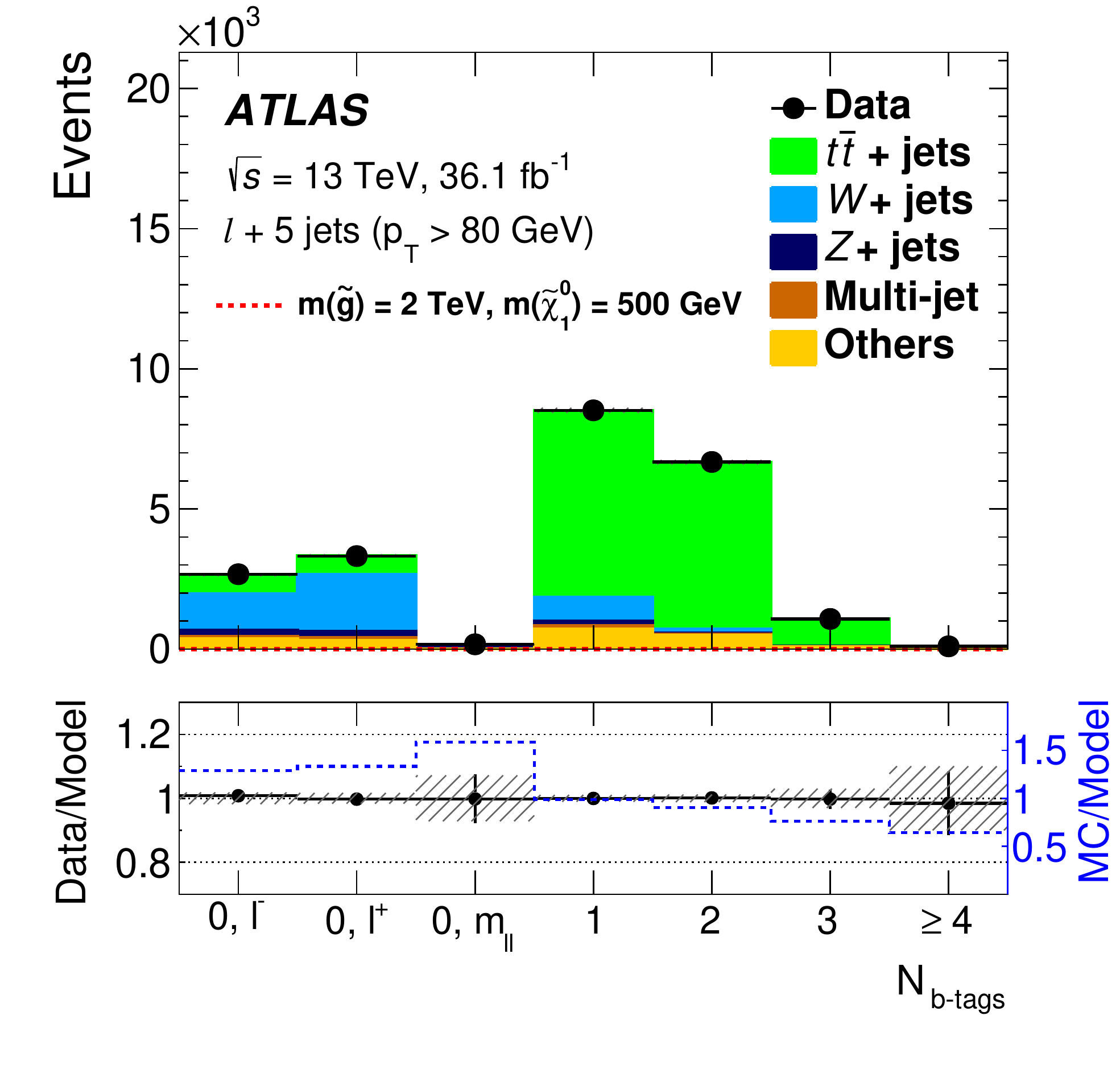}
\includegraphics[width=.43\textwidth]{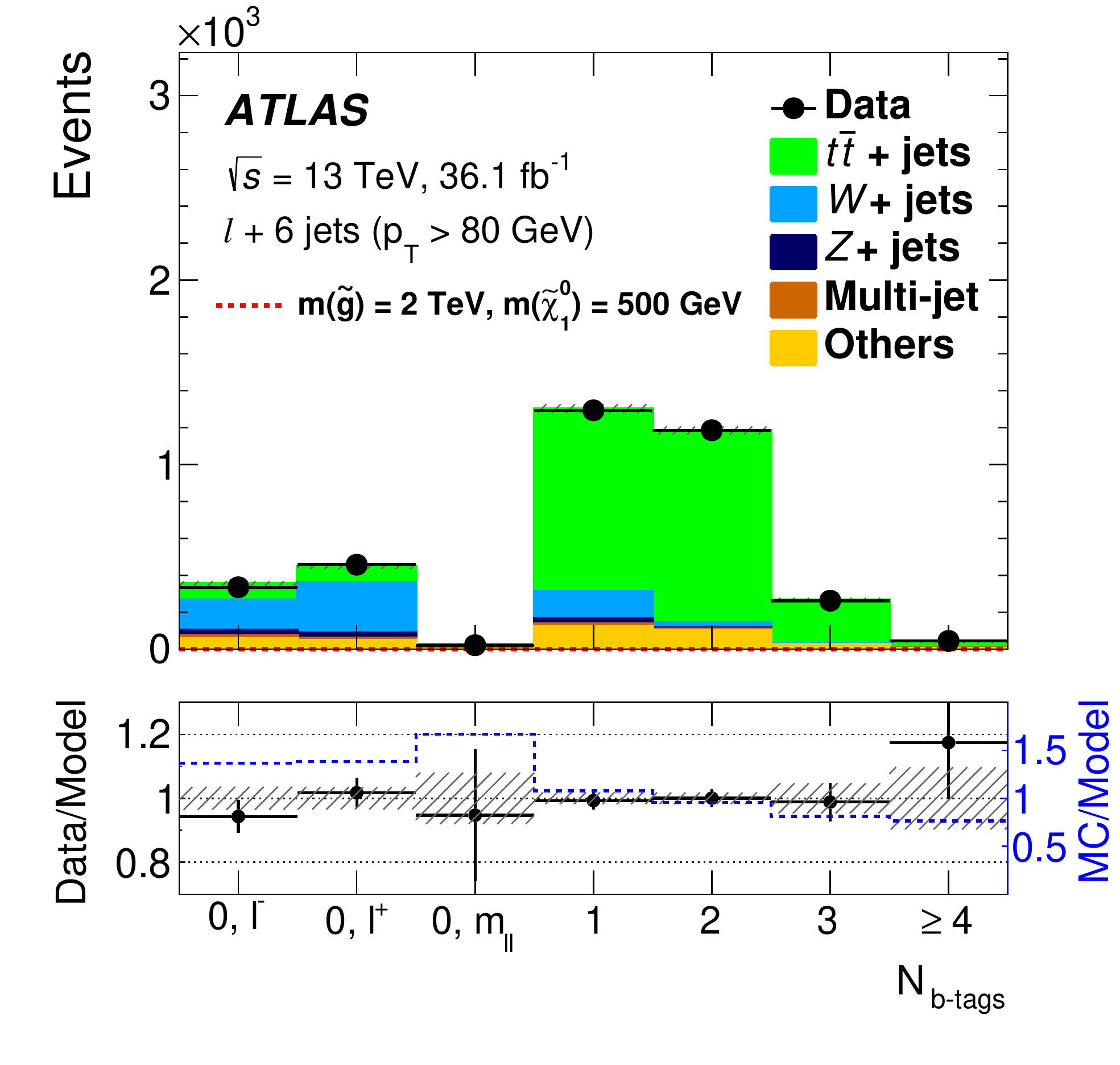}\\
\includegraphics[width=.43\textwidth]{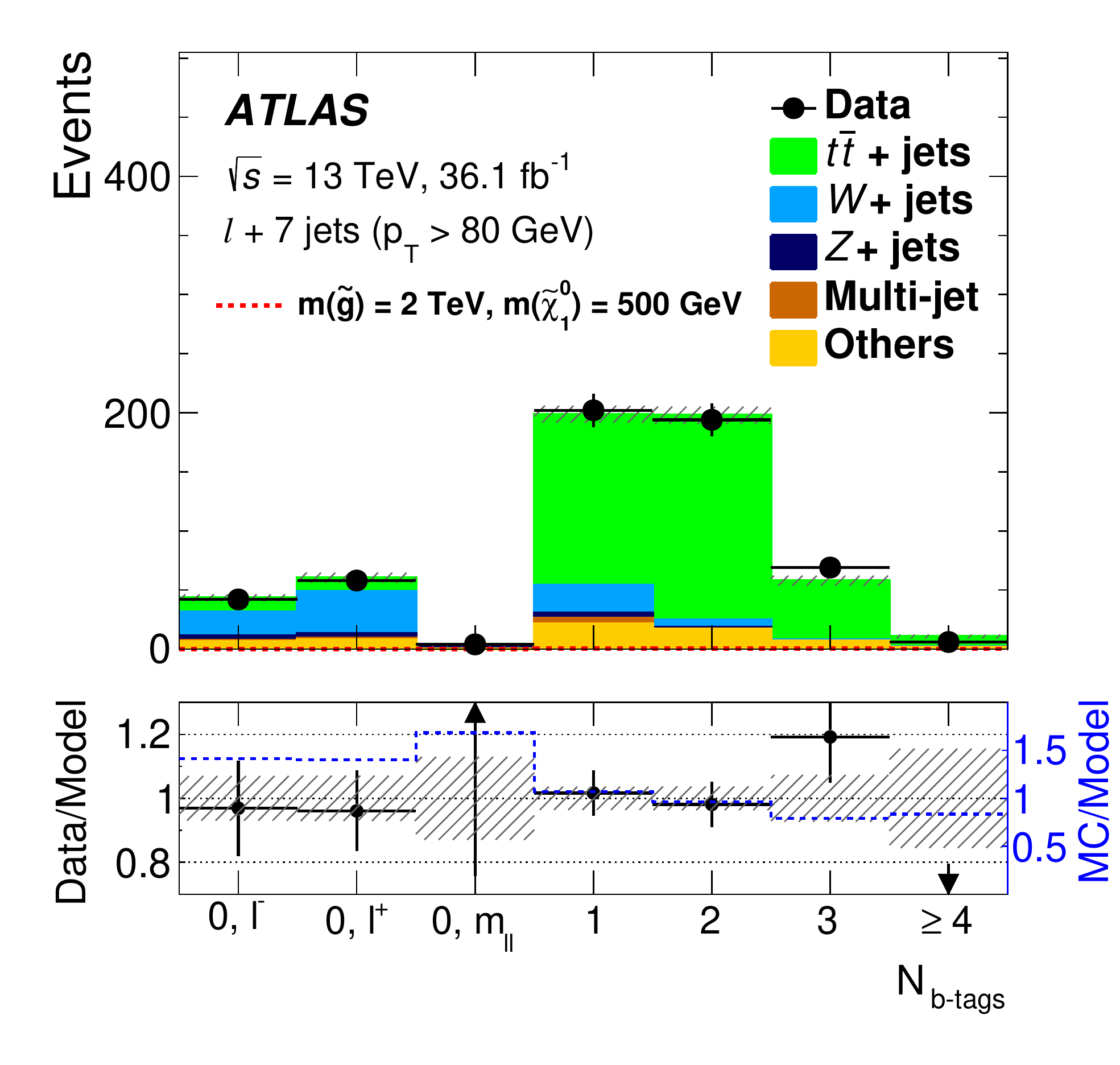}
\includegraphics[width=.43\textwidth]{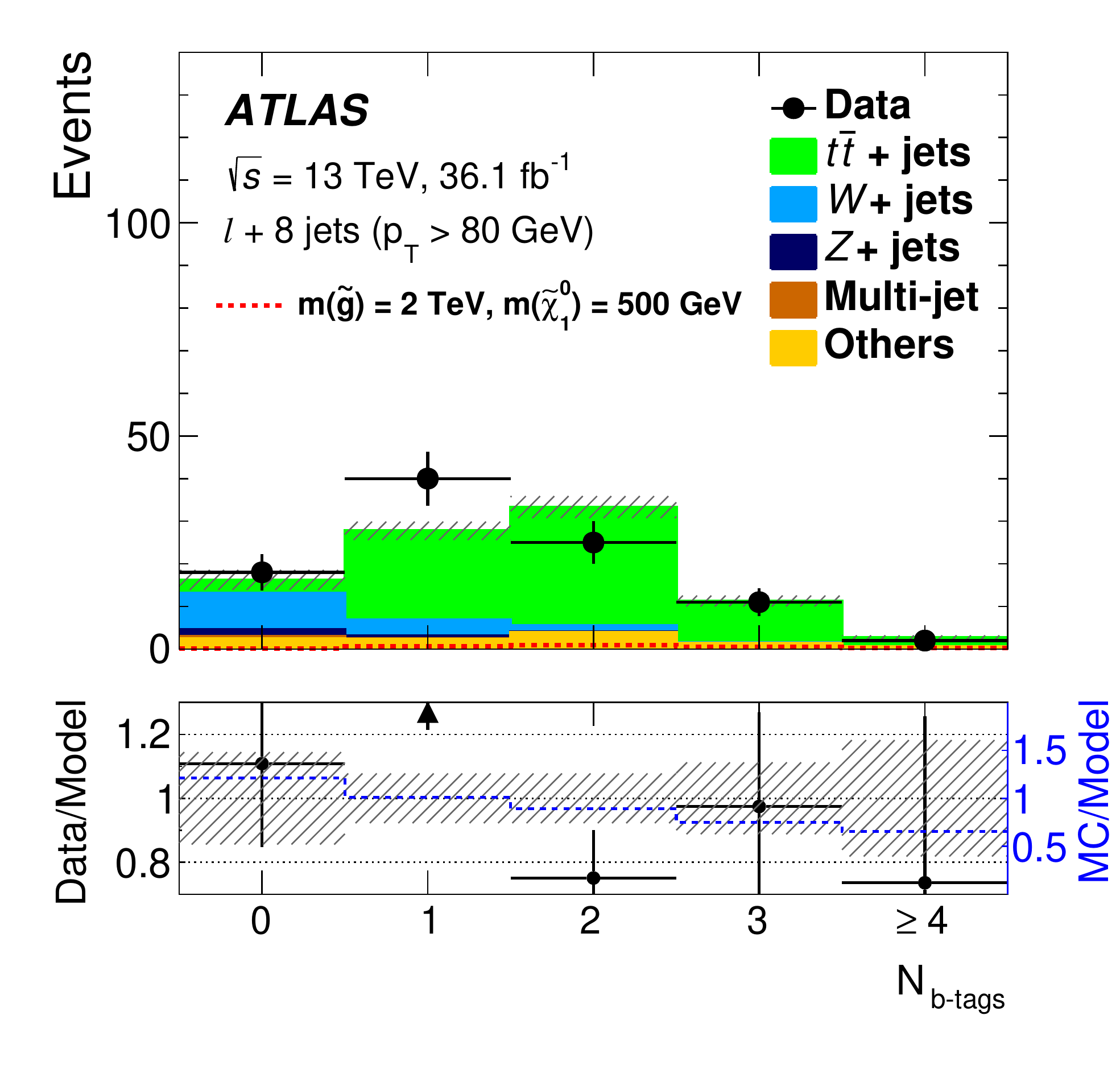}\\
\includegraphics[width=.43\textwidth]{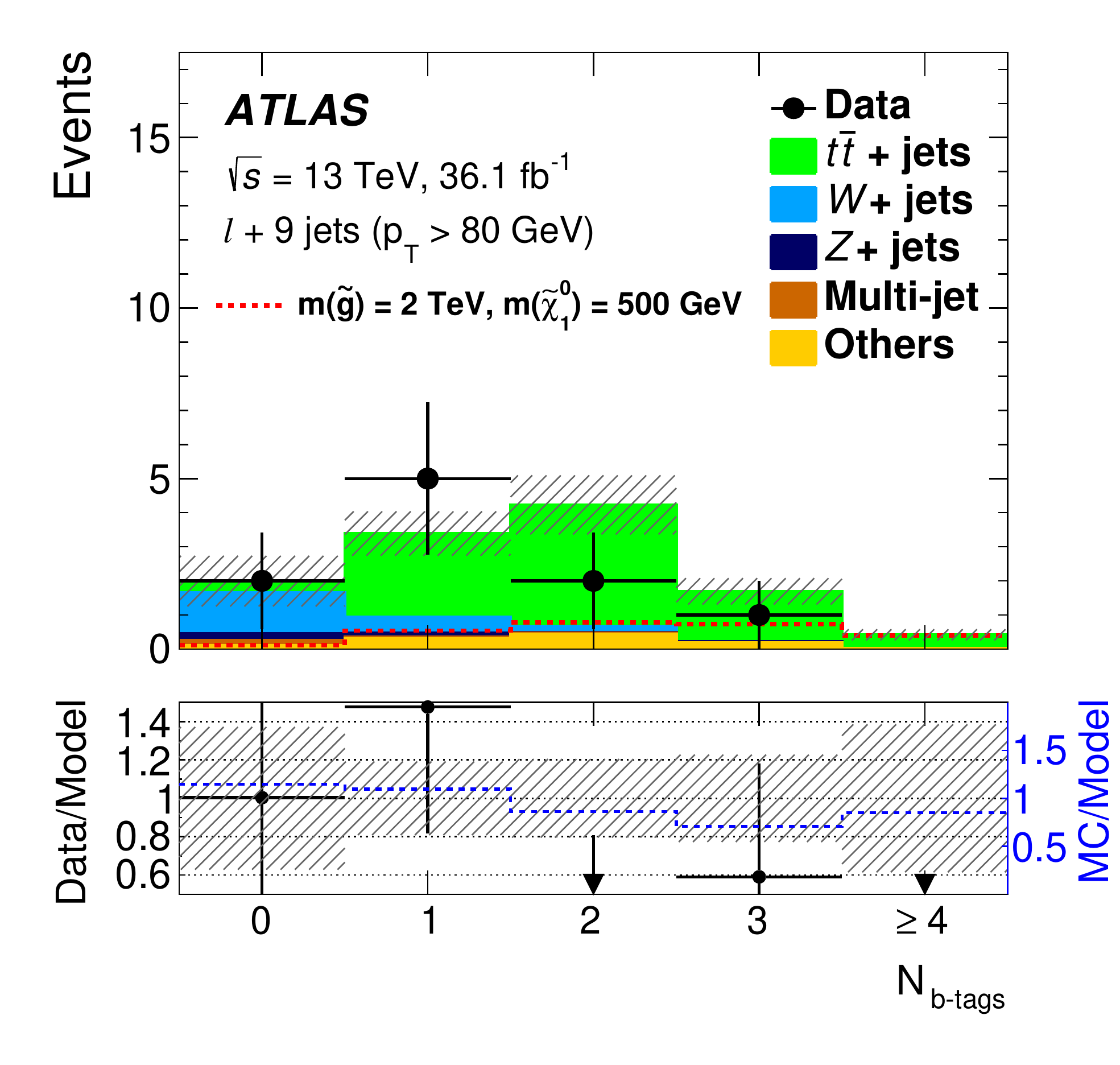}
\includegraphics[width=.43\textwidth]{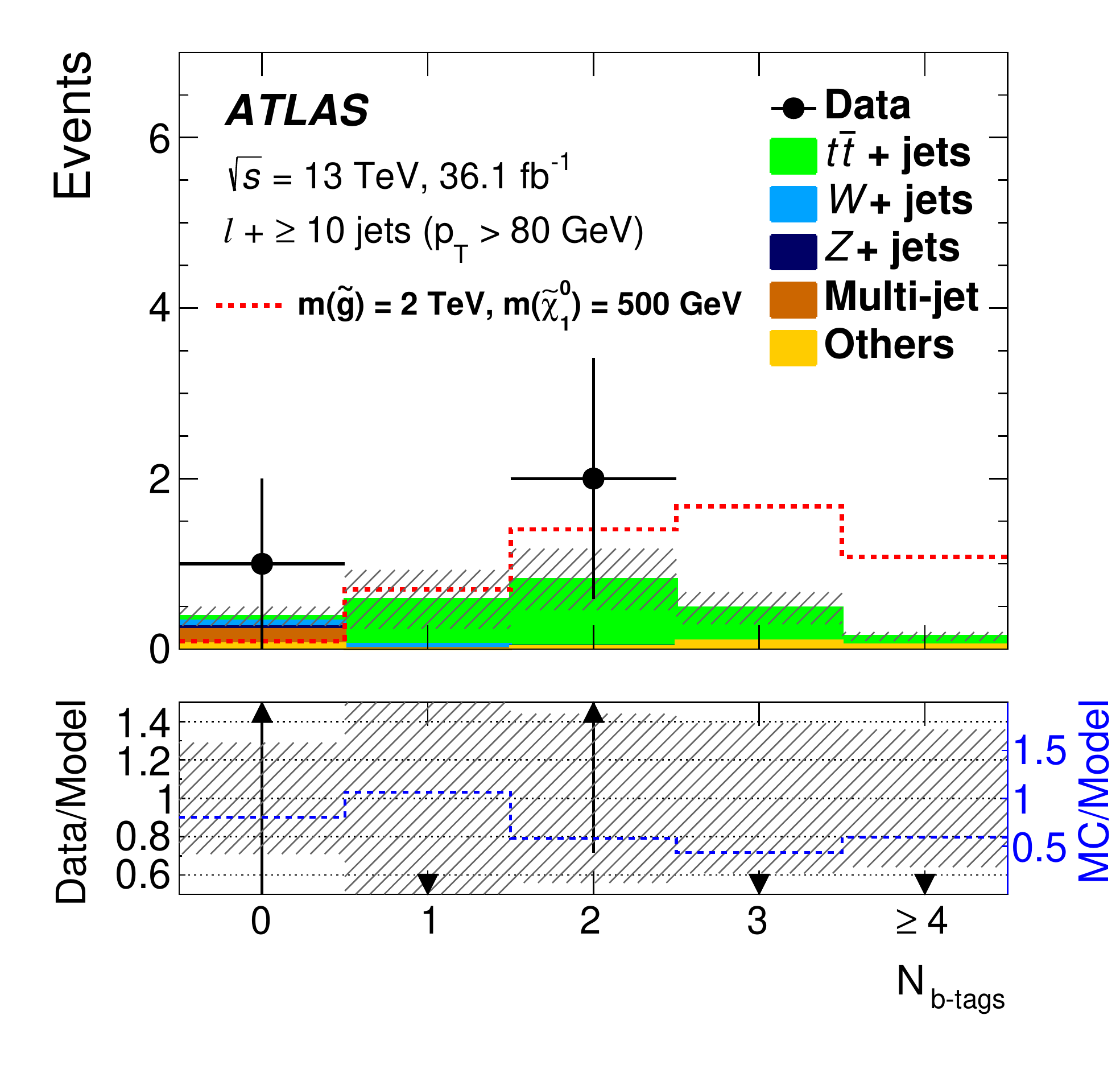}\\
\caption{The expected background and observed data in the different jet and $b$-tag multiplicity bins for the 80~\GeV\ jet \pT\ threshold. The background shown is estimated by including all bins in the
fit. For the five-, six- and seven-jet slices the control regions used to estimate the $W$+jets and $Z$+jets normalizations are also shown (labelled $\ell^-$, $\ell^+$, and $m_{\ell \ell}$). An example
signal for the $\gluino\to t\bar t\ninoone \to  t\bar t u d s$ model with $m_{\gluino}$ = 2000~\GeV\ and $m_{\ninoone}$ = 500~\GeV\ is also overlaid (although its contribution is very small in most of the
jet multiplicity slices shown). 
The bottom panels show the ratio of the observed data to the expected background, as well as the ratio of the prediction from MC simulation to the expected background. 
All uncertainties, which can be correlated across the bins, are included in the error bands (shaded regions).}
\label{fig:postfit_80_77_unblind}
\end{figure}

Figures~\ref{fig:postfit_40_77_unblind}, \ref{fig:postfit_60_77_unblind} and~\ref{fig:postfit_80_77_unblind} show the observed numbers of data events compared to the fitted background model,
for the three jet \pT\ thresholds, respectively. The likelihood fit is configured using the model-dependent set-up where all bins are input to the fit, and fixing the signal-strength parameter to zero. An 
example signal model is also shown to illustrate the separation between the signal and  the background achieved, as well as the level of signal-event leakage into lower $b$-tag and 
jet-multiplicity bins. The bottom panel of each figure shows the background prediction using MC simulation. For high $b$-tag multiplicities ($\ge 3$), the MC simulation strongly 
underestimates the background contributions compared to the data-driven 
background estimation. This effect has been observed before~\cite{CMS-TOP-13-010,Aad:2015gra} and shows that the MC simulations are not able to correctly 
describe final states with high $b$-jet multiplicity. In addition, the MC simulation predicts too many events at low $b$-jet multiplicity, which is likely to be due to a mismodelling of the $W$+jets production 
at high jet multiplicity.  
Since the background prediction from MC simulation does not reflect the expected background contribution, in all cases the expected limit is computed using the background prediction from a fit to all bins
 in the data with no signal component included in the fit model.

\subsection{Model-independent results}
\label{sec:model-indep}

The model-independent results are calculated from the observed number of events, and the expected background in the SRs.
Tables~\ref{tab:yields_SR_40gev},~\ref{tab:yields_SR_60gev}, and~\ref{tab:yields_SR_80gev} show the expected background in the SRs from these fits together with the observed numbers of events for the sets 
of SRs with the 40~\GeV, 60~\GeV\ and 80~\GeV\ jet \pT\ thresholds. In addition, the $p_0$ values are shown, 
which quantify the probability that a background-only experiment results in a fluctuation giving an event yield equal to or larger than the one observed in the data. 
The background estimate describes the observed data in the SRs well, with the largest excesses over the background estimate corresponding to 0.8 standard deviations in SRs 40-3b-11 and 40-3b-12.

\input{tables/discfit_allSRs_40GeV.tex}

\input{tables/discfit_allSRs_60GeV.tex}

\input{tables/discfit_allSRs_80GeV.tex}

Model-independent upper limits at 95\% confidence level (CL) on the number of BSM events, $N_\mathrm{BSM}$, that may contribute to the signal regions, are computed from the observed number of events and the fitted background. Normalizing these 
results by the integrated luminosity $L$ of the data sample, they can be interpreted as upper limits on the visible BSM cross-section $\sigma_{\text{vis}}$, defined as the product 
$\sigma_{\text{prod}}\times A \times\epsilon=N_\mathrm{BSM}/L$ of production cross-section ($\sigma_{\text{prod}}$), acceptance ($A$) and reconstruction efficiency ($\epsilon$). These limits are presented 
in Table~\ref{tab:model_ind_limits_merged}.

\input{tables/Limits_70_merged.tex}

For a hypothetical signal with three or four $b$-jets, the analysis sensitivity is reduced because of the leakage of signal events into lower $b$-tag jet multiplicity bins due to the  
$b$-tagging efficiency of about 78\%, which would bias the normalization of the $t\bar{t}$+jets background. This is partially mitigated by excluding the two-$b$-tag bin in the background determination for the 
highest jet slice probed, and by the constraint on the scaling of the $t\bar{t}$+jets background as a function of jet multiplicity. 

\subsection{Model-dependent results}
\begin{figure}[htb]
\centering
\includegraphics[width=0.49\textwidth]{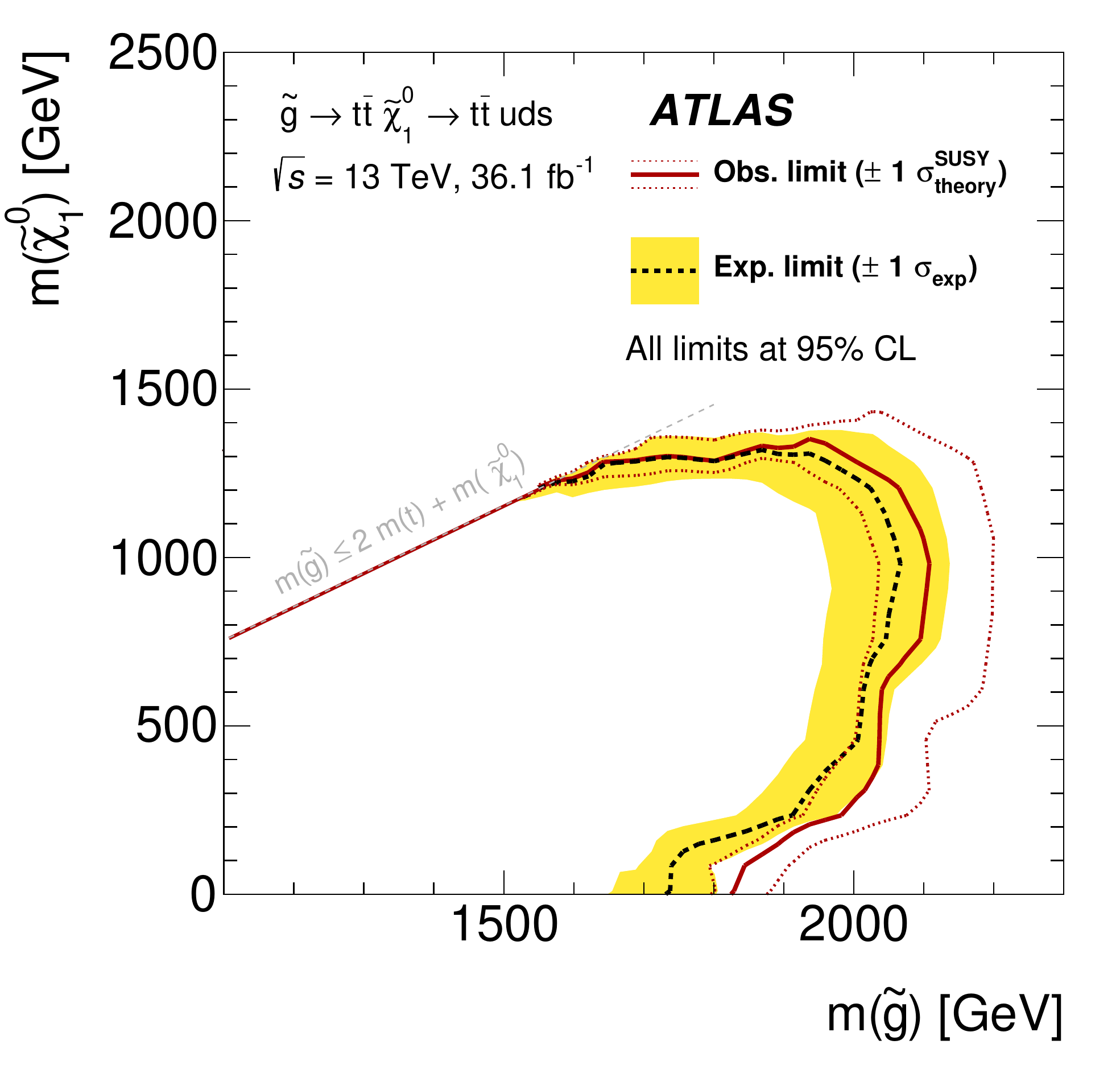}
\includegraphics[width=0.49\textwidth]{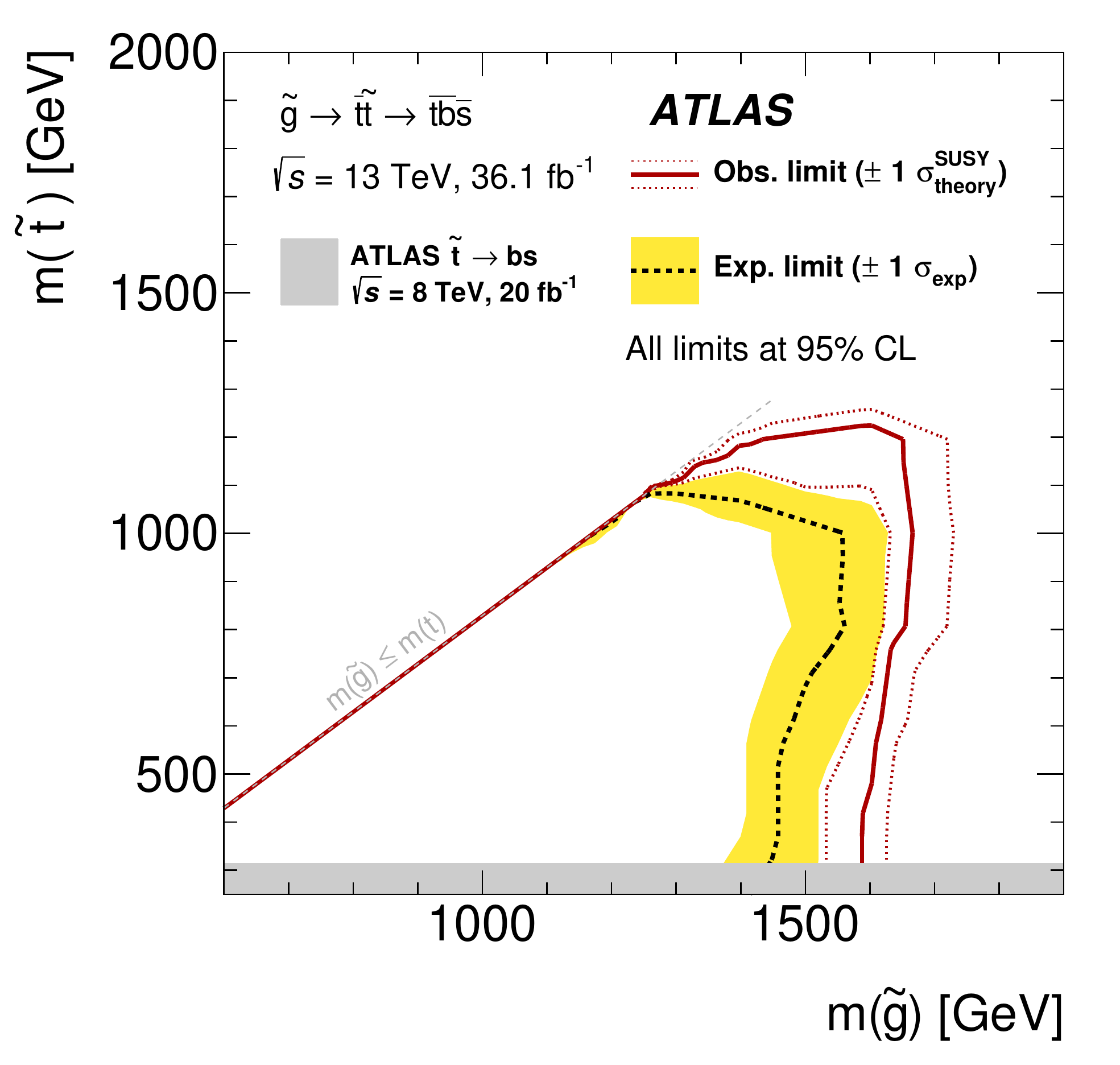}
\includegraphics[width=0.49\textwidth]{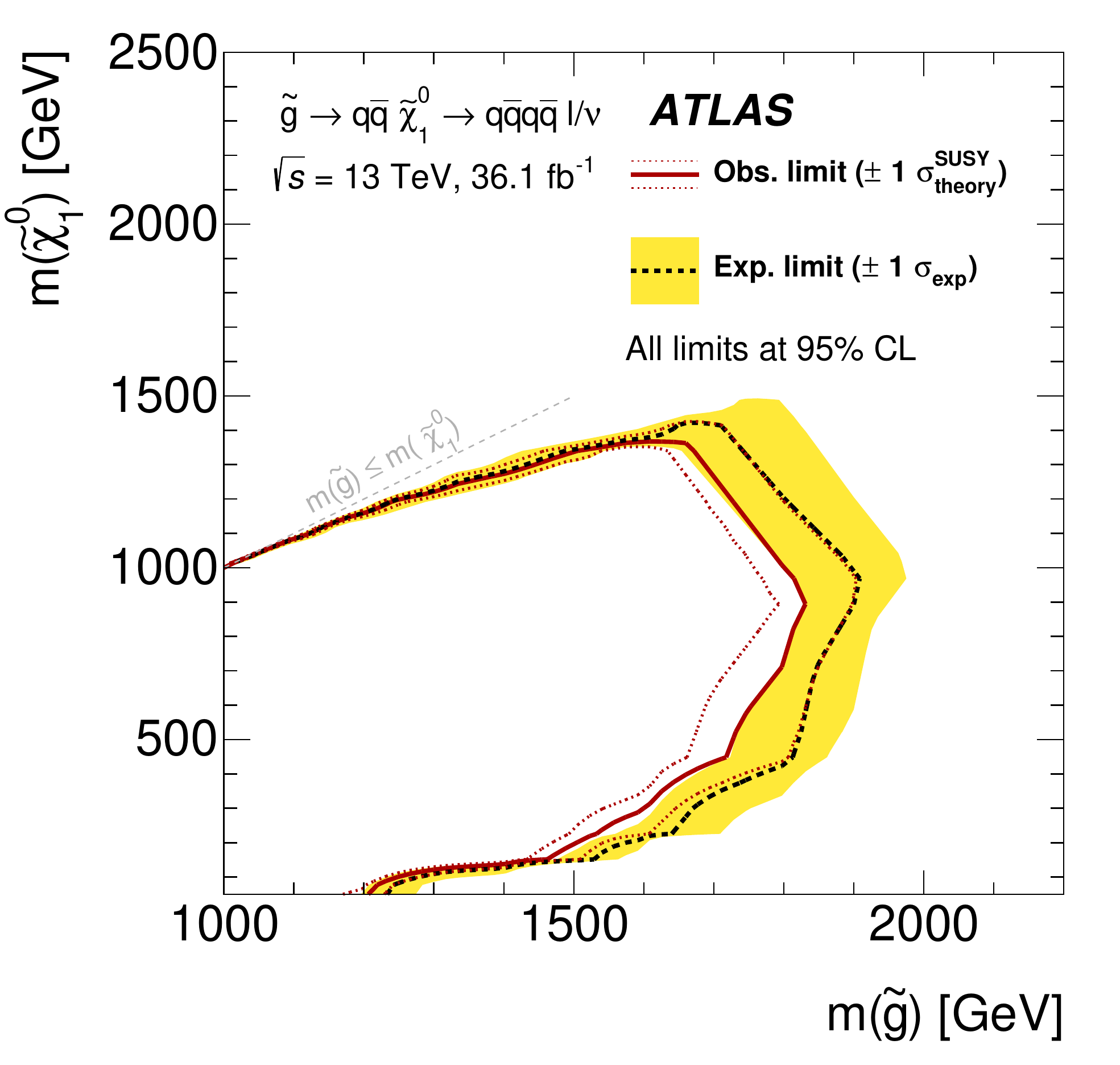}
\caption{
Observed and expected exclusion contours on the \gluino and \ninoone or \stop masses in the context of the RPV SUSY scenarios probed, with simplified mass spectra featuring $\gluino\gluino$ pair production with exclusive decay modes.
The contours of the band around the expected limit are the $\pm$1$\sigma$ variations,  including all uncertainties except theoretical uncertainties in the signal cross-section. The dotted lines around the observed limit illustrate the change in the observed limit as the nominal signal cross-section is scaled up and down by the theoretical uncertainty. All limits are computed at 95\% CL. The diagonal line indicates the kinematic limit for the decays in each specified scenario. For the $\gluino\to \bar t \stop \to \bar t \bar{b} \bar{s}$ model, the limit on the top-squark mass from Ref.~\cite{SUSY-2015-05} is also shown.}
\label{fig:observed_limits}
\end{figure}

\begin{figure}[htb]
\centering
\includegraphics[width=0.65\textwidth]{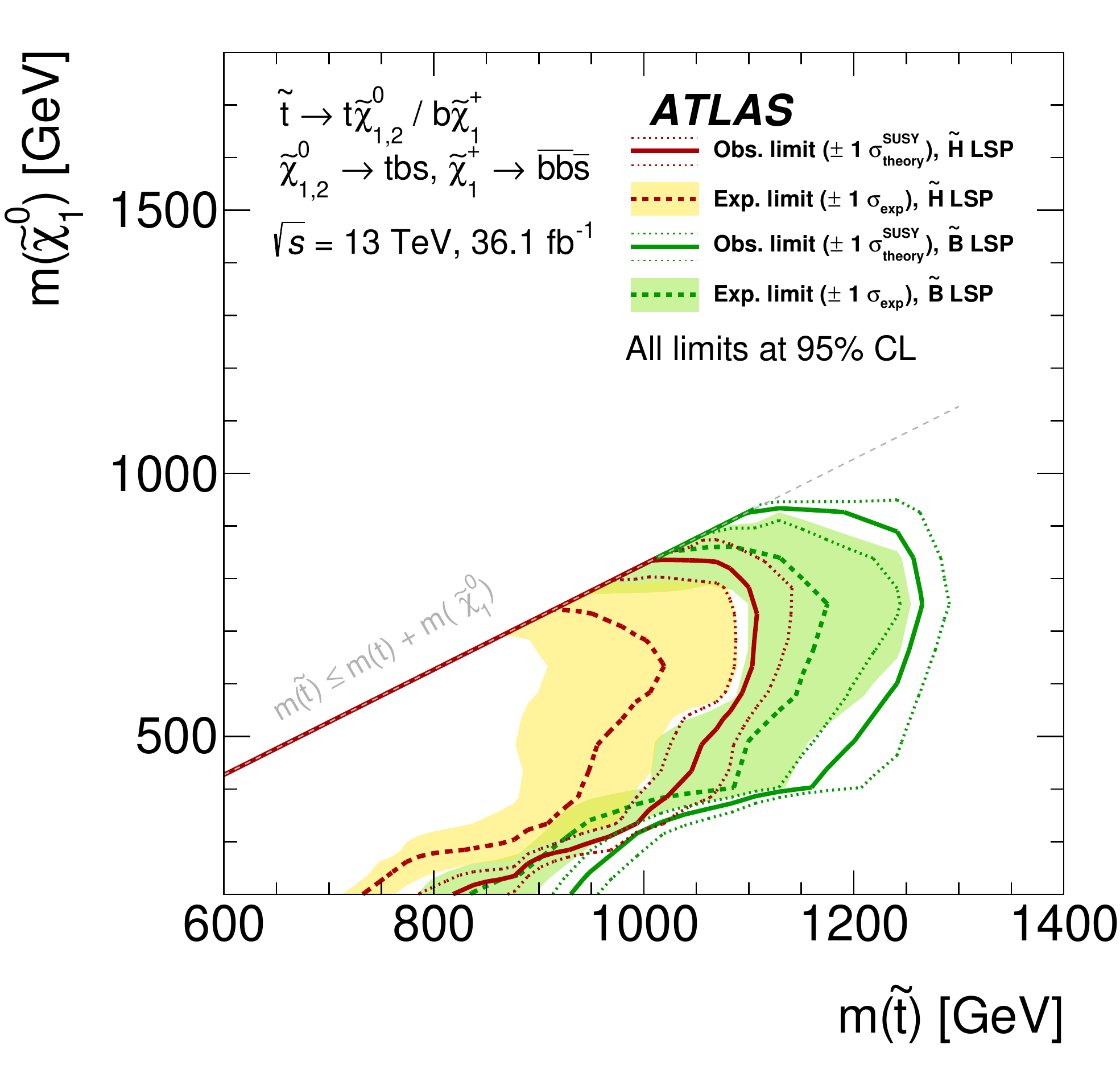}
\caption{%Expected and observed exclusion contours at $95\%$ C.L. in the two benchmark signal models 
Observed and expected exclusion contours on the \stop and \ninoone  masses in the context of top-squark production model with RPV decays of the LSP. Limits are shown in the case of pure bino ($\tilde{B}$) or pure higgsino ($\tilde{H}$) LSPs.  The contours of the band around the expected limit are the $\pm$1$\sigma$ variations,  including all uncertainties except theoretical uncertainties in the signal cross-section. The dotted lines around the observed limit illustrate the change in the observed limit as the nominal signal cross-section is scaled up and down by the theoretical uncertainty. All limits are computed at 95\% CL. The diagonal line indicates the kinematic limit for the decays in the considered scenario.}
\label{fig:observed_limits2}
\end{figure}

For each signal model probed, the fit is configured using the model-dependent set-up, as detailed in Section~\ref{subsec:bkd_validation}. All bins are included in the fit and the expected signal 
contribution in each bin is taken into account. Figure~\ref{fig:observed_limits} shows the observed and expected exclusion limits for the three benchmark signal models featuring gluino pair production, as a 
function of the gluino mass and neutralino or top-squark mass. Figure~\ref{fig:observed_limits2} shows exclusion limits in the top-squark production model where the limit for pure bino and higgsino LSPs are shown separately, taking into account the processes discussed in Section~\ref{sec:signal}.
For the gluino production models, all the probed model points have the best expected sensitivity when using the 80~\GeV\ jet \pT\ threshold, 
whereas for the top-squark production model, the 60~\GeV\ jet \pT\ threshold gives the best expected sensitivity, and these thresholds are used to set the exclusion limits.

In the model with an RPV decay of the \ninoone\ to three light-quark jets, gluino masses up to 2.10~\TeV\ are excluded, with weaker limits for light and heavy \ninoone. For the benchmark 
model with $\gluino\to \bar t \stop$ and $\stop\to \bar{b} \bar{s}$, gluino masses up to 1.65~\TeV\ are excluded. 
In this case, the observed limit is about two standard deviations stronger than the expected limit. This is due to a difference between the observed data and the expected background in the three- and four-$b$-tag bins in the eight-, nine- and ten-jet slices (see Figure~\ref{fig:postfit_80_77_unblind}), which are the  most sensitive bins for this model.
An exclusion limit is also derived for the same model but 
with a virtual top squark (with mass set to 2~\TeV ) where gluinos of mass up to 1.62~\TeV\ are excluded (with an expected exclusion up to 1.50~\TeV ). 
The analysis excludes gluinos with masses up to 1.80~\TeV\ in the $\gluino \to q \bar q \ninoone \to q \bar q q \bar q \ell / \nu$ model.

For the top-squark pair production model, top-squark masses up to 1.10~\TeV\ and 1.25~\TeV\ are excluded for higgsino and bino LSPs respectively. 
There is greater sensitivity in the case of the bino LSP because the lepton and jet multiplicities are higher than in the higgsino LSP scenario.\footnote{In the bino case, every top-squark decay produces a top quark whereas for higgsino LSPs top quarks are produced in only about half of the cases.} 

Typical acceptance times efficiency ($A \times \epsilon$) values for the relevant SR for each of the benchmark signal models are:
\begin{itemize}
\item 8\% for the $\gluino\to t\bar t\ninoone \to  t\bar t u d s$ model for the 80-3b-10 SR,
\item 3\% for the $\gluino\to \bar t \stop \to \bar t \bar{b} \bar{s}$ model for the 80-3b-8 SR,
\item 13\% for the $\gluino \to q \bar q \ninoone \to q \bar q q \bar q \ell / \nu$ model for the 80-0b-8 SR,
\item 2\% (6\%) for the top-squark production model with a higgsino (bino) LSP for the 60-3b-10 SR.
\end{itemize}
These values correspond to the case where the produced SUSY particle is close to the exclusion limit, and for intermediate LSP masses. In general, the acceptance falls for light or heavy LSPs as some of the produced jets or leptons become softer.

\subsection{Limits on four-top-quark production}

The analysis is also used to search for SM four-top-quark production. In this case, the small contribution to the background from four-top-quark production is removed, and a model-dependent fit is 
carried out with the four-top-quark simulated sample used as the signal. The best expected sensitivity is achieved with the 60~\GeV\ jet \pT\ threshold, which leads to a cross-section upper limit at 95\% CL 
on the four-top-quark signal of 60~fb (whereas 84~fb is expected), which is 6.5 
times the SM cross-section for this process
.\footnote{No uncertainty in the theoretical modelling of the four-top-quark process is included when setting the cross-section limit, although uncertainties related to the $b$-tagging, jet and lepton reconstruction are taken into account.}

%% file: tables/discfit_allSRs_40GeV.tex
\begin{table}[!tbh]
\begin{center}

\begin{tabular}{| c | c | c | c | c | c | c |}
\hline
 & \multicolumn{2}{c |}{$ \geq 10 $ jets}& \multicolumn{2}{c |}{$ \geq 11 $ jets}& \multicolumn{2}{c |}{$ \geq 12 $ jets} \\
\hline
Process & $0$ b  & $\geq 3$ b & $0$ b  & $\geq 3$ b & $0$ b  & $\geq 3$ b  \\
\hline
 $t\bar{t}$+jets  & $14.3\pm2.9\hd$  & $53\pm6\hd$  & $3.0\pm0.7$  & $10.5\pm1.8\hd$  & $0.58\pm0.20$  & $1.9\pm0.6$  \\
 $W$+jets  & $7\pm4$  & $0.22\pm0.08$  & $0.9\pm0.9$  & $0.04\pm0.03$  & $0.1\pm0.1$  & $<0.01$  \\
 Others  & $1.9\pm0.6$  & $6.3\pm1.8$  & $0.19\pm0.06$  & $1.7\pm0.6$  & $0.05\pm0.02$  & $0.57\pm0.20$  \\
 $Z$+jets  & $1.7\pm0.9$  & $0.10\pm0.03$  & $0.25\pm0.22$  & $0.02\pm0.01$  & $0.03\pm0.04$  & $<0.01$  \\
 Multi-jet  & $1.3\pm0.7$  & $0.48\pm0.20$  & $0.15\pm0.08$  & $0.27\pm0.12$  & $0.12\pm0.07$  & $<0.01$  \\
\hline
 Total Bkd. & $26\pm4\hd$ & $60\pm6\hd$ & $4.5\pm1.0$ & $12.6\pm1.9\hd$ & $0.87\pm0.23$ & $2.5\pm0.7$ \\
 Data  & 23  & 61  & 5  & 16  & 0  & 4  \\
\hline
 $p_0$ ($\sigma$)  & 0.5 (0)  & 0.46 (0.1)  & 0.42 (0.2)  & 0.21 (0.8)  & 0.5 (0)  & 0.21 (0.8)  \\
\hline

    \end{tabular}
    \caption{Fitted background yields in the different $b$-tag multiplicity bins for jet $\pt > 40~\GeV$ in the different signal regions. The parameters of the model are determined in a fit to a reduced set of bins, corresponding to the model-independent fit  discussed in the text. The individual background uncertainties can be larger than the total uncertainty due to correlations between parameters. The $p_0$ value quantifies the probability that a background-only experiment results in a fluctuation giving an event yield equal to or larger than the one observed in the data, and is capped at 0.5.}
    \label{tab:yields_SR_40gev}
    \end{center}
    \end{table}

%% file: tables/discfit_allSRs_60GeV.tex
\begin{table}[!tbh]
\begin{center}

\begin{tabular}{| c | c | c | c | c | c | c |}
\hline
 & \multicolumn{2}{c |}{$ \geq 8 $ jets}& \multicolumn{2}{c |}{$ \geq 9 $ jets}& \multicolumn{2}{c |}{$ \geq 10 $ jets} \\
\hline
Process & $0$ b  & $\geq 3$ b & $0$ b  & $\geq 3$ b & $0$ b  & $\geq 3$ b  \\
\hline
 $t\bar{t}$+jets  & $26\pm11$  & $88\pm17$  & $4.0\pm1.9$  & $20.7\pm3.4\hd$  & $0.56\pm0.33$  & $4.1\pm1.7$  \\
 $W$+jets  & $42\pm9\hd$  & $1.18\pm0.31$  & $7.1\pm2.4$  & $0.24\pm0.09$  & $1.2\pm1.1$  & $0.02\pm0.02$  \\
 Others  & $11\pm4\hd$  & $12\pm4\hd$  & $2.2\pm0.8$  & $3.0\pm0.9$  & $0.34\pm0.11$  & $0.77\pm0.26$  \\
 $Z$+jets  & $8.0\pm1.3$  & $0.32\pm0.04$  & $1.3\pm0.4$  & $0.09\pm0.02$  & $0.23\pm0.19$  & $0.02\pm0.02$  \\
 Multi-jet  & $3.0\pm1.5$  & $0.50\pm0.24$  & $0.56\pm0.27$  & $0.27\pm0.13$  & $0.32\pm0.15$  & $<0.01$  \\
\hline
 Total Bkd. & $90\pm9\hd$ & $102\pm17\hd$ & $15.1\pm2.5\hd$ & $24.4\pm3.3\hd$ & $2.7\pm1.2$ & $4.9\pm1.7$ \\
 Data  & 91  & 102  & 18  & 15  & 2  & 2  \\
\hline
 $p_0$ ($\sigma$)  & 0.46 (0.1)  & 0.5 (0)  & 0.27 (0.6)  & 0.5 (0)  & 0.5 (0)  & 0.5 (0)  \\
\hline

    \end{tabular}
    \caption{Fitted background yields in the different $b$-tag multiplicity bins for jet $\pt > 60~\GeV$ in the different signal regions. The parameters of the model are determined in a fit to a reduced set of bins, corresponding to the model-independent fit  discussed in the text. The individual background uncertainties can be larger than the total uncertainty due to correlations between parameters. The $p_0$ value quantifies the probability that a background-only experiment results in a fluctuation giving an event yield equal to or larger than the one observed in the data, and is capped at 0.5.}
    \label{tab:yields_SR_60gev}
    \end{center}
    \end{table}

%% file: tables/discfit_allSRs_80GeV.tex
\begin{table}[!tbh]
\begin{center}

\begin{tabular}{| c | c | c | c | c | c | c |}
\hline
 & \multicolumn{2}{c |}{$ \geq 8 $ jets}& \multicolumn{2}{c |}{$ \geq 9 $ jets}& \multicolumn{2}{c |}{$ \geq 10 $ jets} \\
\hline
Process & $0$ b  & $\geq 3$ b & $0$ b  & $\geq 3$ b & $0$ b  & $\geq 3$ b  \\
\hline
 $t\bar{t}$+jets  & $4.0\pm1.7$  & $15.7\pm2.3\hd$  & $0.44\pm0.21$  & $2.7\pm0.7$  & $0.06\pm0.04$  & $0.38\pm0.27$  \\
 $W$+jets  & $9.0\pm2.9$  & $0.18\pm0.07$  & $1.2\pm0.7$  & $0.02\pm0.02$  & $0.05\pm0.07$  & $<0.01$  \\
 Others  & $2.3\pm0.9$  & $2.4\pm0.7$  & $0.16\pm0.05$  & $0.45\pm0.15$  & $0.06\pm0.03$  & $0.14\pm0.05$  \\
 $Z$+jets  & $1.7\pm0.5$  & $0.06\pm0.02$  & $0.23\pm0.14$  & $0.03\pm0.01$  & $0.02\pm0.03$  & $<0.01$  \\
 Multi-jet  & $0.8\pm0.4$  & $<0.01$  & $0.30\pm0.15$  & $<0.01$  & $0.16\pm0.08$  & $<0.01$  \\
\hline
 Total Bkd. & $17.8\pm2.9\hd$ & $18.4\pm2.2\hd$ & $2.3\pm0.9$ & $3.2\pm0.7$ & $0.35\pm0.13$ & $0.52\pm0.27$ \\
 Data  & 21  & 14  & 3  & 1  & 1  & 0  \\
\hline
 $p_0$ ($\sigma$)  & 0.27 (0.6)  & 0.5 (0)  & 0.34 (0.4)  & 0.5 (0)  & 0.18 (0.9)  & 0.5 (0)  \\
\hline

    \end{tabular}
    \caption{Fitted background yields in the different $b$-tag multiplicity bins for jet $\pt > 80~\GeV$ in the different signal regions. The parameters of the model are determined in a fit to a reduced set of bins, corresponding to the model-independent fit  discussed in the text. The individual background uncertainties can be larger than the total uncertainty due to correlations between parameters. The $p_0$ value quantifies the probability that a background-only experiment results in a fluctuation giving an event yield equal to or larger than the one observed in the data, and is capped at 0.5.}
    \label{tab:yields_SR_80gev}
    \end{center}
    \end{table}

%% file: tables/Limits_70_merged.tex
\begin{table}[ptbh]
\begin{center}
  \def\arraystretch{1.2}
\begin{tabular}{| l  c | c | c | c | c | c | c |}
\hline
\multicolumn{2}{| c |}{Jet multiplicity} & $0$ b obs. [fb]  &  $0$ b exp. [fb]  &  $\geq 3$ b obs. [fb]  &  $\geq 3$ b exp. [fb]   \\
\hline
$\geq10$ jets&($\pt > 40~\GeV$) & $0.32$  & $0.36^{+0.16}_{-0.10}$
 & $0.57$  & $0.54^{+0.24}_{-0.15}$ \\
$\geq11$ jets&($\pt > 40~\GeV$) & $0.17$  & $0.16^{+0.08}_{-0.05}$
 & $0.33$  & $0.25^{+0.12}_{-0.07}$ \\
$\geq12$ jets&($\pt > 40~\GeV$) & $0.08$  & $0.09^{+0.05}_{-0.01}$
 & $0.17$  & $0.13^{+0.07}_{-0.04}$ \\
\hline
$\geq8$ jets&($\pt > 60~\GeV$)  & $0.73$  & $0.71^{+0.27}_{-0.20}$
 & $1.02$  & $1.03^{+0.39}_{-0.29}$ \\
$\geq9$ jets&($\pt > 60~\GeV$)  & $0.35$  & $0.28^{+0.12}_{-0.08}$
 & $0.19$  & $0.32^{+0.15}_{-0.09}$ \\
$\geq10$ jets&($\pt > 60~\GeV$)  & $0.12$  & $0.14^{+0.07}_{-0.04}$
 & $0.11$  & $0.15^{+0.08}_{-0.04}$ \\
\hline
$\geq8$ jets&($\pt > 80~\GeV$)& $0.38$  & $0.31^{+0.14}_{-0.09}$
 & $0.21$  & $0.28^{+0.13}_{-0.08}$ \\
$\geq9$ jets&($\pt > 80~\GeV$)& $0.15$  & $0.13^{+0.07}_{-0.04}$
 & $0.09$  & $0.13^{+0.07}_{-0.04}$ \\
$\geq10$ jets&($\pt > 80~\GeV$)&  $0.10$  & $0.08^{+0.04}_{-0.00}$
 & $0.08$  & $0.08^{+0.04}_{-0.00}$ \\
\hline

    \end{tabular}
    \caption{Observed and expected 95\% CL model-independent upper limits on the product of cross-section, acceptance and efficiency (in fb) for each signal region. The limits are determined by fitting the background model in a reduced set of bins as described in the text.}
    \label{tab:model_ind_limits_merged}
    \end{center}
    \end{table}

%% file: conclusions.tex
\section{Conclusion}
\label{sec:conclusion}

A search for beyond the Standard Model physics in events with an isolated lepton (electron or muon), high jet multiplicity and no, or many, $b$-tagged jets is presented. 
Unlike many previous searches in similar final states, no requirement on the missing transverse momentum in the event is applied. A novel data-driven technique is used to estimate the dominant backgrounds from $t\bar{t}$+jets and $W/Z$+jets production. 
The analysis is performed with proton--proton collision data at $\sqrt{s}=13~\TeV$ collected in 2015 and 2016 with the ATLAS detector at the Large Hadron Collider 
corresponding to an integrated luminosity of 36.1~fb$^{-1}$. 
With no significant excess over the Standard Model expectation observed, results are interpreted in the framework of simplified models featuring gluino or top-squark pair production in $R$-parity-violating supersymmetry scenarios. 
In a benchmark model with $\gluino\to t\bar t\ninoone \to  t\bar t u d s$, gluino masses up to 2.10~\TeV\ are excluded at 95\% confidence level. 
In a model with $\gluino\to \bar t \stop$ and $\stop\to \bar b \bar s$, gluino masses up to 1.65~\TeV\ are excluded, whereas in a model with $\gluino \to q \bar q \ninoone \to q \bar q q \bar q \ell / \nu$, gluino masses up to 1.80~\TeV\ are excluded. A model with direct top-squark production and $R$-parity-violating decays of higgsino or bino LSPs excludes top squarks with masses up to 1.10~\TeV\ and 1.25~\TeV\ respectively. 
These results improve the previously existing limits for the gluino production models considered, whereas they represent the first limits for the top squark production model. 
In addition, an upper limit of 60~fb is set on the cross-section of Standard Model four-top-quark production, improving on the previous strongest limit of 69~fb~\cite{Sirunyan:2017tep}. 
Finally, model-independent limits are set on the contribution of new phenomena to the signal-region yields.

%% file: Acknowledgements.tex
% Acknowledgements for papers with collision data
% Version 6-Mar-2017

% Standard acknowledgements start here
%----------------------------------------------
We thank CERN for the very successful operation of the LHC, as well as the
support staff from our institutions without whom ATLAS could not be
operated efficiently.

We acknowledge the support of ANPCyT, Argentina; YerPhI, Armenia; ARC, Australia; BMWFW and FWF, Austria; ANAS, Azerbaijan; SSTC, Belarus; CNPq and FAPESP, Brazil; NSERC, NRC and CFI, Canada; CERN; CONICYT, Chile; CAS, MOST and NSFC, China; COLCIENCIAS, Colombia; MSMT CR, MPO CR and VSC CR, Czech Republic; DNRF and DNSRC, Denmark; IN2P3-CNRS, CEA-DSM/IRFU, France; SRNSF, Georgia; BMBF, HGF, and MPG, Germany; GSRT, Greece; RGC, Hong Kong SAR, China; ISF, I-CORE and Benoziyo Center, Israel; INFN, Italy; MEXT and JSPS, Japan; CNRST, Morocco; NWO, Netherlands; RCN, Norway; MNiSW and NCN, Poland; FCT, Portugal; MNE/IFA, Romania; MES of Russia and NRC KI, Russian Federation; JINR; MESTD, Serbia; MSSR, Slovakia; ARRS and MIZ\v{S}, Slovenia; DST/NRF, South Africa; MINECO, Spain; SRC and Wallenberg Foundation, Sweden; SERI, SNSF and Cantons of Bern and Geneva, Switzerland; MOST, Taiwan; TAEK, Turkey; STFC, United Kingdom; DOE and NSF, United States of America. In addition, individual groups and members have received support from BCKDF, the Canada Council, CANARIE, CRC, Compute Canada, FQRNT, and the Ontario Innovation Trust, Canada; EPLANET, ERC, ERDF, FP7, Horizon 2020 and Marie Sk{\l}odowska-Curie Actions, European Union; Investissements d'Avenir Labex and Idex, ANR, R{\'e}gion Auvergne and Fondation Partager le Savoir, France; DFG and AvH Foundation, Germany; Herakleitos, Thales and Aristeia programmes co-financed by EU-ESF and the Greek NSRF; BSF, GIF and Minerva, Israel; BRF, Norway; CERCA Programme Generalitat de Catalunya, Generalitat Valenciana, Spain; the Royal Society and Leverhulme Trust, United Kingdom.

The crucial computing support from all WLCG partners is acknowledged gratefully, in particular from CERN, the ATLAS Tier-1 facilities at TRIUMF (Canada), NDGF (Denmark, Norway, Sweden), CC-IN2P3 (France), KIT/GridKA (Germany), INFN-CNAF (Italy), NL-T1 (Netherlands), PIC (Spain), ASGC (Taiwan), RAL (UK) and BNL (USA), the Tier-2 facilities worldwide and large non-WLCG resource providers. Major contributors of computing resources are listed in Ref.~\cite{ATL-GEN-PUB-2016-002}.
%----------------------------------------------

%% file: atlas_authlist.tex
% ATLAS Collaboration author list
% Data extracted on 27-Mar-2017 for paper reference SUSY-2016-11
\begin{flushleft}
{\Large The ATLAS Collaboration}

\bigskip

M.~Aaboud$^\textrm{\scriptsize 137d}$,
G.~Aad$^\textrm{\scriptsize 88}$,
B.~Abbott$^\textrm{\scriptsize 115}$,
O.~Abdinov$^\textrm{\scriptsize 12}$$^{,*}$,
B.~Abeloos$^\textrm{\scriptsize 119}$,
S.H.~Abidi$^\textrm{\scriptsize 161}$,
O.S.~AbouZeid$^\textrm{\scriptsize 139}$,
N.L.~Abraham$^\textrm{\scriptsize 151}$,
H.~Abramowicz$^\textrm{\scriptsize 155}$,
H.~Abreu$^\textrm{\scriptsize 154}$
% ...
\bigskip
\\
% ...
$^{12}$ Institute of Physics, Azerbaijan Academy of Sciences, Baku, Azerbaijan\\
% ...
$^{88}$ CPPM, Aix-Marseille Universit{\'e} and CNRS/IN2P3, Marseille, France\\
% ...
$^{115}$ Homer L. Dodge Department of Physics and Astronomy, University of Oklahoma, Norman OK, United States of America\\
% ...
$^{119}$ LAL, Univ. Paris-Sud, CNRS/IN2P3, Universit{\'e} Paris-Saclay, Orsay, France\\
% ...
$^{137}$ $^{(a)}$ Facult{\'e} des Sciences Ain Chock, R{\'e}seau Universitaire de Physique des Hautes Energies - Universit{\'e} Hassan II, Casablanca; $^{(b)}$ Centre National de l'Energie des Sciences Techniques Nucleaires, Rabat; $^{(c)}$ Facult{\'e} des Sciences Semlalia, Universit{\'e} Cadi Ayyad, LPHEA-Marrakech; $^{(d)}$ Facult{\'e} des Sciences, Universit{\'e} Mohamed Premier and LPTPM, Oujda; $^{(e)}$ Facult{\'e} des sciences, Universit{\'e} Mohammed V, Rabat, Morocco\\
% ...
$^{139}$ Santa Cruz Institute for Particle Physics, University of California Santa Cruz, Santa Cruz CA, United States of America\\
% ...
$^{151}$ Department of Physics and Astronomy, University of Sussex, Brighton, United Kingdom\\
% ...
$^{154}$ Department of Physics, Technion: Israel Institute of Technology, Haifa, Israel\\
$^{155}$ Raymond and Beverly Sackler School of Physics and Astronomy, Tel Aviv University, Tel Aviv, Israel\\
% ...
$^{161}$ Department of Physics, University of Toronto, Toronto ON, Canada\\
% ...
$^{a}$ Also at Department of Physics, King's College London, London, United Kingdom\\
$^{b}$ Also at Institute of Physics, Azerbaijan Academy of Sciences, Baku, Azerbaijan\\
$^{c}$ Also at Novosibirsk State University, Novosibirsk, Russia\\
$^{d}$ Also at TRIUMF, Vancouver BC, Canada\\
$^{e}$ Also at Department of Physics {\&} Astronomy, University of Louisville, Louisville, KY, United States of America\\
% ...
$^{*}$ Deceased
\end{flushleft}